\documentclass[journal]{IEEEtran}
\usepackage{amsmath,graphicx}
\usepackage[linesnumbered,ruled]{algorithm2e}
\usepackage{multirow}
\usepackage{hhline}
\usepackage{cite}
\usepackage{pifont}
\usepackage{xcolor}
\newcommand{\tabincell}[2]{\begin{tabular}{@{}#1@{}}#2\end{tabular}}
\newcommand{\xmark}{\ding{55}}
\newcommand{\cmark}{\ding{51}}

\ifCLASSINFOpdf
\else
\fi
\begin{document}
%
\title{Bayesian Learning of LF-MMI Trained Time Delay Neural Networks for Speech Recognition}
%
%
%

\author{Shoukang Hu, Xurong Xie, Shansong Liu, Jianwei Yu, Zi Ye, Mengzhe Geng, \\Xunying Liu,~\IEEEmembership{Member,~IEEE}, Helen Meng,~\IEEEmembership{Fellow,~IEEE}
  \thanks{Shoukang Hu (e-mail: skhu@se.cuhk.edu.hk), Shansong Liu, Jianwei Yu, Zi Ye, Mengzhe Geng, Xunying Liu (e-mail: xyliu@se.cuhk.edu.hk), Helen Meng (e-mail: hmmeng@se.cuhk.edu.hk) are with the Department of Systems Engineering and Engineering Management, The Chinese University of Hong Kong, Hong Kong SAR, China. Xurong Xie (e-mail: xr.xie@siat.ac.cn) is with Shenzhen Institutes of Advanced Technology, Chinese Academy of Sciences, Shenzhen, China. (Corresponding author: Xunying Liu.)}
}

\markboth{IEEE/ACM Transactions on Audio, Speech and Language Processing}%
{Bare Demo of IEEEtran.cls for IEEE Journals}
%



\maketitle

\begin{abstract}
Discriminative training techniques define state-of-the-art performance for automatic speech recognition systems. However, they are inherently prone to overfitting, leading to poor generalization performance when using limited training data. In order to address this issue, this paper presents a full Bayesian framework to account for model uncertainty in sequence discriminative training of factored TDNN acoustic models. Several Bayesian learning based TDNN variant systems are proposed to model the uncertainty over weight parameters and choices of hidden activation functions, or the hidden layer outputs. Efficient variational inference approaches using a few as one single parameter sample ensure their computational cost in both training and evaluation time comparable to that of the baseline TDNN systems. Statistically significant word error rate (WER) reductions of 0.4\%-1.8\% absolute (5\%-11\% relative) were obtained over a state-of-the-art 900 hour speed perturbed Switchboard corpus trained baseline LF-MMI factored TDNN system using multiple regularization methods including F-smoothing, L2 norm penalty, natural gradient, model averaging and dropout, in addition to i-Vector plus learning hidden unit contribution (LHUC) based speaker adaptation and RNNLM rescoring. The efﬁcacy of the proposed Bayesian techniques is further demonstrated in a comparison against the state-of-the-art performance obtained on the same task using the most recent hybrid and end-to-end systems reported in the literature. Consistent performance improvements were also obtained on a 450 hour HKUST conversational Mandarin telephone speech recognition task. On a third cross domain adaptation task requiring rapidly porting a 1000 hour LibriSpeech data trained system to a small DementiaBank elderly speech corpus, the proposed Bayesian TDNN LF-MMI systems outperformed the baseline system using direct weight fine-tuning by up to 2.5\% absolute WER reduction.
\end{abstract}


%
\IEEEpeerreviewmaketitle

\section{Introduction}
\IEEEPARstart{T}{here} has been a long history of using discriminative training techniques to improve the performance of automatic speech recognition (ASR) systems. In current neural network based systems, these discriminative training methods define the state-of-the-art performance. From the previous generation of Gaussian mixture model based Hidden Markov Models (HMMs) ASR systems~\cite{bahl1986maximum, povey2008boosted, povey2005discriminative, kaiser2000novel, gibson2006hypothesis, povey2007evaluation} to the current systems using a hybrid HMM deep neural network (DNN) architecture~\cite{kingsbury2009lattice, vesely2013sequence, su2013error, povey2016purely, sak2014sequence, michel2020frame}, performance improvements obtained over the conventional cross-entropy (CE) trained systems have been widely reported. Although in recent years there has been a significant trend of moving from hybrid HMM-DNN system architectures to all neural end-to-end (E2E) modelling paradigm represented by listen, attend and spell (LAS)~\cite{chan2016listen}, connectionist temporal classification (CTC)~\cite{graves2006connectionist}, RNN transducers (RNN-T)\cite{graves2012sequence} and neural transformers~\cite{wang2020transformer}, state-of-the-art hybrid HMM-DNN systems featuring sequence discriminative training techniques, for example, maximum mutual information (MMI) criterion~\cite{bahl1986maximum, hadian2018end, michel2019comparison} trained factored time delay neural networks (TDNNs)~\cite{waibel1989consonant, peddinti2015time, povey2018semi, povey2016purely}, remain highly competitive against end-to-end approaches to date~\cite{hadian2018end, luscher2019rwth, zhou2020rwth, li2020comparison}. 

Since discriminative training methods were first introduced to the earlier generation of GMM-HMM based speech recognition systems~\cite{bahl1986maximum, valtchev1997mmie, woodland2002large}, they have been long known to be prone to overfitting when using limited training data and a sparse representation of the modelling confusion over possible erroneous recognition hypotheses. In the context of deep neural network based ASR systems, this overfitting issue also presents~\cite{benesty2008automatic}, for example, when using smaller sized and shallower lattices to train systems with a very large number of HMM state targets~\cite{su2013error}. Such issue is further aggravated by the use of stochastic gradient based optimization techniques that operate sequentially in a batch mode on smaller subsets of data randomly drawn from the complete training data collection.

In order to address the above issue, several categories of techniques have been developed in recent years to improve the generalization performance of discriminative training for DNN based ASR systems. Drawing inspirations from the earlier regularization techniques used in the discriminative training of GMM-HMM systems~\cite{povey2002minimum}, the first category of methods attempts to alleviate the problem by optimizing the interpolated error cost between a sequence level discriminative training criterion, for example, MMI, and the conventional CE cost, as in F-smoothing~\cite{su2013error}.  Motivated by the data intensive nature of deep learning techniques, the second category of techniques reduce the risk of overfitting using data augmentation methods. By expanding the limited training data using, for example, speed perturbation~\cite{ko2015audio}, spectral deformation~\cite{park2019specaugment}, simulation of noisy and reverberated speech~\cite{ko2017study}, the coverage of the augmented training data and the resulting speech recognition systems' generalization performance can be improved. The third category of methods address the overfitting issue by modifying the optimization algorithm. These include the incorporation of an additional L2 norm term into the original discriminative error cost function~\cite{vesely2013sequence}. Second-order methods represented by Hessian-free optimization~\cite{kingsbury2012scalable, sainath2013accelerating, chung2017parallel} and the use of natural gradient~\cite{yang1998complexity, roux2008topmoumoute, DBLP:journals/corr/PoveyZK14} have been investigated. Model parameter averaging~\cite{DBLP:journals/corr/PoveyZK14} over different batch intervals at the end of a training epoch can also be used, for example, employed in the Kaldi toolkit as a standard regularisation technique. Weight noise adds noise directly to the network parameters to improve generalization~\cite{murray1994enhanced, braun2019parameter}. Finally, dropout, a simple and effective approach used in deep learning models to avoid over-fitting~\cite{srivastava2014dropout} can also be adopted. However, it lacks of a mathematically well defined framework to model the underlying DNN ASR systems’ uncertainty arising from the use of limited data, when compared with Bayesian neural networks~\cite{DBLP:conf/icml/GalG16}. Most of the above regularization techniques are currently used in the Kaldi implementation of lattice-free maximum mutual information (LF-MMI) sequence training of factored TDNN systems~\cite{povey2016purely}.

This paper presents a mathematically well grounded, full Bayesian framework to account for model uncertainty in sequence discriminative training of factored TDNN acoustic models. In contrast to conventional Bayesian neural networks marginalizing over the CE error cost, an integration of the sequence level MMI criterion is used. Modelling uncertainty is addressed using three full Bayesian approaches. Bayesian TDNN systems~\cite{lam2018gaussian, hu2019bayesian, hu2019lf, xie2019blhuc, li2020bayesian} are used to model uncertainty over the weight parameters. Gaussian Process TDNN systems are further introduced to consider both the uncertainty associated with the weight parameters, as well as that over the choice of hidden activation functions. Variational TDNN systems are proposed to consider the uncertainty over the hidden layer outputs. Efficient variational inference approaches developed for all the above Bayesian TDNN systems using a very small number of samples (as low as one) ensure their computational cost in both training and evaluation time comparable to that of the baseline TDNN systems. A theoretical connection is further drawn between full Bayesian inference and dropout by re-formulating the latter as a special case of Bayesian TDNN systems. 

Experiments conducted on a state-of-the-art 900 hour speed perturbed Switchboard corpus trained baseline LF-MMI factored TDNN system featuring multiple built-in regularization methods including F-smoothing~\cite{su2013error}, L2 norm~\cite{vesely2013sequence}, natural gradient~\cite{yang1998complexity, roux2008topmoumoute, DBLP:journals/corr/PoveyZK14}, model averaging~\cite{DBLP:journals/corr/PoveyZK14} and dropout~\cite{srivastava2014dropout}, as well as i-Vector~\cite{dehak2010front, madikeri2016implementation} and learning hidden unit contribution (LHUC)~\cite{swietojanski2014learning} speaker adaptation suggests the proposed Bayesian TDNN, Gaussain Process TDNN and variational TDNN systems consistently outperform the baseline systems by a statistically significant margin of 0.4\%-1.8\% absolute (5\%-11\% relative) reduction in word error rate over the NIST Hub5’00, RT02 and RT03 sets. Similar consistent performance improvements were also obtained after the recurrent neural network language model rescoring, as well as on a 450 hour (with  speed perturbation) HKUST conversational Mandarin telephone speech recognition task. The efficacy of the proposed Bayesian estimation techniques is further demonstrated on a cross domain adaptation task. A 1000 Hour LibriSpeech corpus trained LF-MMI TDNN system is rapidly domain adapted to a highly challenging elderly speech recognition corpus based on a 10 hour Dementia Bank Pitt database. Consistent performance improvements of 1.1\% absolute WER reduction over LF-MMI baseline TDNN systems using direct weight fine-tuning were obtained.

The main contributions of this paper are summarized below: 

1) This paper presents a first use of a mathematically well grounded, full Bayesian framework to account for model uncertainty in sequence discriminative training of factored TDNN acoustic models. A systematic overview and comparison over different full Bayesian TDNN learning variants is given. In contrast, only limited previous research on Bayesian neural network learning techniques was conducted for language modelling~\cite{chien2016bayesian}. More recently, a Bayesian learning framework was used to account for model uncertainty in sequence discriminative training of factored TDNN acoustic models in our preliminary research~\cite{hu2019bayesian, hu2019lf}. Stochastic noise injection to model parameters~\cite{murray1994enhanced} was also exploited to improve the generalization performance of E2E ASR systems~\cite{braun2019parameter,tuske2020single}.

2) Efficient variational inference approaches developed for all the above Bayesian TDNN systems using a very small number of samples (as low as one) ensures their computational cost comparable to that of the baseline systems. The generic nature of the proposed methods also allows them to be extended to other end-to-end approaches to address similar modelling uncertainty issues during system development. 

3) Significant performance improvements on multiple data sets were obtained over baseline LF-MMI factored TDNN systems constructed using a large ensemble of built-in regularization methods including F-smoothing, L2 norm penalty, natural gradient, model averaging and dropout.

4) This paper further presents the earliest work on full Bayesian learning driven rapid domain adaptation of LF-MMI TDNN based ASR systems. In contrast to the previous research based on transfer learning~\cite{ghahremani2017investigation}, the proposed Bayesian domain adaptation technique provides an alternative useful approach to the problem of under-resourced speech recognition system development.

The rest of this paper is organized as follows. Section 2 introduces a full Bayesian learning framework for several neural network model variants that account for the uncertainty over the weight parameters, and the choice of activation functions or the hidden layer outputs. These include Bayesian neural networks (BNNs), Gaussian Process neural networks (GPNNs) and Variational neural networks (VNNs). Time delay neural networks (TDNNs) are presented in Section 3. Section 4 discusses the Bayesian estimation of TDNNs. Section 5 shows the experiments and results. Finally, the conclusions are drawn in Section 6. 

\section{Bayesian Leaning Based Neural Network}
In this section, we introduce several forms of Bayesian learning based neural networks presented in this paper, including Bayesian Neural Networks (BNNs), Gaussian Process Neural Networks (GPNNs) and Variational Neural Networks (VNNs). 



\subsection{Bayesian Neural Network}
\label{bnn}
In conventional neural networks using fixed-point parameter estimates, the uncertainty associated with the prediction is hard to quantify. Bayesian neural networks (BNNs) offer a formalism to understand and quantify the uncertainty by using the posterior distribution to model the parameter uncertainty in the predictive distribution~\cite{mackay1992practical, neal2012bayesian, bishop2006pattern, graves2011practical, pmlr-v37-blundell15}. To make predictions for the observations of test utterance $\mathbf{O}_r^{*}$, we average over all the parameter values, weighted by their posterior probability. 
\begin{equation}
\label{eq: bayesian_likelihood}
p(\mathbf{H}_r^{*}|\mathbf{O}_r^{*},\mathbf{D}) \\
=\int p(\mathbf{H}_r^{*}|\mathbf{O}_r^{*},\mathbf{w}) p({\mathbf{w}}|\mathbf{D})d\mathbf{w}
\end{equation}
where $\mathbf{H}_r^{*}$ is the predicted word-sequence for utterance $r$, $\mathbf{w}$ is the activation parameter, $p(\mathbf{w}\!\mid\!\mathbf{D})$ denotes the posterior distribution to be learned from training data $\mathbf{D}  =\{\mathbf{H}_r,\mathbf{O}_r\}$, $\mathbf{O}_r$ is the sequence of observation for utterance $r$ and $\mathbf{H}_r$ is the reference word transcription for utterance $r$.

If all subsequent layers $l+1,\cdots,L$ are removed, the expected hidden node output $h_{i}^{(l)}$ of the $i$-th node in the $l$-th layer is marginalized over different parameter estimates.
\begin{equation}
\label{eq: bayesian}
h_{i}^{(l)}=\int\phi\left({{\mathbf{w}_{i}^{(l)}}}\bullet \mathbf{h}^{(l-1)}\right) p({\mathbf{w}_{i}^{(l)}}\mid\mathbf{D})d\mathbf{w}_{i}^{(l)}
\end{equation}
where $\mathbf{h}^{(l-1)}$ is the input vector fed into the $l$-th hidden layer (the output from the previous layer $l-1$), $p(\mathbf{w}_{i}^{(l)}\!\mid\!\mathbf{D})$ denotes the node dependent activation parameter posterior distribution, $\phi\left(\cdot\right)$ is the activation function and $\bullet$ denotes the dot product.

\vspace{-2mm}
\subsection{Gaussian Process Neural Network}
\label{gpnn}
Gaussian Processes (GPs)~\cite{rasmussen2006gaussian} are powerful nonparametric distributions over continuous functions that are used in probabilistic modelling for many machine learning applications including regression and classification tasks and beyond. A function modelled using Gaussian process is represented as 
\begin{equation}
\label{eq:gp_standard}
\begin{aligned}
    f(\mathbf x)&\sim\mathcal{GP}\left(m(\mathbf x),k(\mathbf x,\mathbf x')\right)
\end{aligned}
\end{equation}
where $\mathbf x,\mathbf x'$ are arbitrary inputs, $m(\cdot)$ is the mean function and $k(\cdot,\cdot)$ is the kernel function.

The above formulation is known as the kernel space view of GP models~\cite{rasmussen2006gaussian}. The associated computational complexity over the kernel covariance function during inference is determined by the size of the training data, and therefore impractical to be directly applied in the large-scale tasks, for example, speech recognition systems that often use tens of millions of frame samples or more in training. An alternative and computationally more tractable form of GP models uses basis function interpolation (see Chapter 2 of~\cite{rasmussen2006gaussian}), leads to the following weight space view of GP,
\begin{equation}
\label{eq:gp_weight_space}
\begin{aligned}
    f(\mathbf x)=\pmb{\lambda}^{T} \bullet \phi(\mathbf x)=\sum_{m}\lambda^{m}\phi^{m}(\mathbf x)
\end{aligned}
\end{equation}
where $k(\cdot,\cdot)=\phi(\cdot)^{T}\phi(\cdot)$, $\pmb{\lambda}\sim \mathcal N(\cdot, \cdot)$ represents amplitudes of different basis functions $\phi^{m}(\mathbf x)$ in $\phi(\cdot)$.

The connection between neural networks and Gaussian Processes has also been extensively studied. Based on MacKay’s work on the Bayesian neural network~\cite{mackay1992practical}, Neal~\cite{neal1996priors} proved that single-hidden-layer Bayesian neural networks of infinite width are equivalent to Gaussian Processes~\cite{rasmussen2006gaussian}. Hazan and Jaakkola~\cite{hazan2015steps} and later Lee~\cite{lee2017deep} proposed the use of GP kernels to approximate infinitely wide deep neural networks. In deep Gaussian processes (DGPs)~\cite{damianou2013deep} models deep belief neural network layers were replaced by Gaussian Processes.

The form of traditional Bayesian neural networks introduced earlier in Sec.~\ref{bnn} only considers the uncertainty associated with weight parameters, but not the network structural configurations. For example, the choice over the hidden activation functions can be learned using a simple output level interpolation of commonly used basis activation functions, i.e., Sigmoid, Tanh, ReLU as the following.
\begin{equation}
\label{eq: gpnn0}
\setlength{\abovedisplayskip}{4pt}
\setlength{\belowdisplayskip}{4pt}
\begin{aligned}
h_{i}^{(l)}
&= \sum_m\lambda_{i}^{(l,m)}\phi_{m}\left({\mathbf{w}_{i}^{(l,m)}}\bullet\mathbf{h}^{(l-1)}\right)
\end{aligned} 
\end{equation}
where $\lambda_{i}^{(l,m)}$ is the $m$-th basis activation coefficient and $\phi_{m}$ is the $m$-th basis activation function.

Within a more general framework of Gaussian Process neural networks (GPNNs), not only the weight parameters inside the activation functions can be treated as random variables, the additional uncertainty over the basis coefficients can also be considered. The prediction is rewritten as
\begin{equation}
\setlength{\abovedisplayskip}{4pt}
\setlength{\belowdisplayskip}{4pt}
\begin{aligned}
\label{eq: gp_likelihood}
&p(\mathbf{H}_r^{*}|\mathbf{O}_r^{*},\mathbf{D})=\int\int p(\mathbf{H}_r^{*}|\mathbf{O}_r^{*},\mathbf{w},\pmb{\lambda}) p({\mathbf{w}}|\mathbf{D}) p({\pmb{\lambda}}|\mathbf{D})d\mathbf{w}d\pmb{\lambda}
\end{aligned}
\end{equation}
where $p(\pmb{\lambda}\!\mid\!\mathbf{D})$ and $p(\mathbf{w}\!\mid\!\mathbf{D})$ denote the basis activation coefficient and parameter posterior distributions respectively. We assume these two variables are independent. The general form of Gaussian Process Neural Network can be further simplified into four special cases in Tab.~\ref{uncertainty_table} by considering different uncertainty modelling combinations (marginalization over both $\mathbf{w}$ and $\pmb{\lambda}$ or only one of them).

Similarly, the expected hidden node output $h_{i}^{(l)}$ in GPNN can be modified into the integration of both the weight parameters and basis coefficients in Eqn. (\ref{eq: gp_general}).
\begin{equation}
\label{eq: gp_general}
\setlength{\abovedisplayskip}{4pt}
\setlength{\belowdisplayskip}{4pt}
\begin{aligned}
h_{i}^{(l)}
&= \sum_m\int\int\lambda_{i}^{(l,m)}\phi_{m}\left({\mathbf{w}_{i}^{(l,m)}}\bullet\mathbf{h}^{(l-1)}\right)\\
&p(\mathbf{w}_{i}^{(l,m)}\!\mid\!\mathbf{D})p(\lambda_{i}^{(l,m)}\!\mid\!\mathbf{D})d\mathbf{w}_{i}^{(l,m)}d\lambda_{i}^{(l,m)}
\end{aligned} 
\end{equation}
where $\mathbf{h}^{(l-1)}$ is the input vector fed into the $l$-th hidden layer, $p(\lambda_{i}^{(l,m)}\!\mid\!\mathbf{D})$ and $p(\mathbf{w}_{i}^{(l,m)}\!\mid\!\mathbf{D})$ denote the basis activation coefficient and parameter posterior distributions. 
\vspace{-3mm}
\subsection{Variational Neural Network}
In contrast to the BNN and GPNN models presented in Sec.~\ref{bnn} and Sec.~\ref{gpnn}, when modelling the uncertainty associated with hidden layer outputs, variational neural networks (VNNs)~\cite{DBLP:journals/corr/KingmaW13, chung2015recurrent, chien2017variational, yu2019comparative} can be used. Instead of modelling the uncertainty over the weight parameters inside the activation functions in BNNs or assuming additional uncertainty over the activation basis coefficients in GPNNs, variational neural networks introduce a latent variable $\mathbf{Z}$ to encode the uncertainty associated with the hidden layer outputs, and in turn the final predictive distribution.
\begin{equation}
\begin{aligned}
\label{eq: variational_likelihood}
&p(\mathbf{H}_r^{*}|\mathbf{O}_r^{*},\mathbf{D})=\int p(\mathbf{H}_r^{*}|\mathbf{O}_r^{*},\mathbf{Z}_r^{*}) p({\mathbf{Z}_r^{*}}|\mathbf{O}_r^{*}, \mathbf{D})d\mathbf{Z}_r^{*}
\end{aligned}
\end{equation}
where $p(\mathbf{Z}_r^{*}\!\mid\!\mathbf{O}_r^{*},\mathbf{D})$ denotes the latent variable posterior distribution to be learned from the training data $\mathbf{D}$.

Similarly, the expected hidden node output is calculated as in Eqn. (\ref{eq: variation}).
\begin{equation}
\label{eq: variation}
\mathbf{h}^{(l)}\!=\!\int\!\phi\left({[{\mathbf{Z}_r^{*}}^{(l-1)}, \mathbf{h}^{(l-1)}]}\right)\! p({{\mathbf{Z}_r^{*}}^{(l-1)}\!\mid\!\mathbf{O}_r^{*},\mathbf{D}})d{\mathbf{Z}_r^{*}}^{(l-1)}
\end{equation}
where $p({\mathbf{Z}_r^{*}}^{(l-1)}\!\mid\!\mathbf{O}_r^{*},\mathbf{D})$ denotes the latent variable posterior distribution of layer $l$, $\phi\left(\cdot\right)$ is the activation function. Note that majority of systems only consider one layer to apply the variational distribution.


\section{Time Delay Neural Network}
Time delay neural networks (TDNNs)~\cite{waibel1989phoneme, waibel1989consonant, peddinti2015time, povey2018semi, povey2016purely, hadian2018end} based hybrid HMM-DNN acoustic models in recent years defined state-of-the-art speech recognition performance over a wide range of tasks, due to their strong power in modelling long range temporal dependencies in speech. In particular, the recently proposed factored TDNN systems~\cite{povey2018semi} featuring lattice-free MMI sequence discriminative training~\cite{povey2016purely} remain highly competitive against all neural end-to-end approaches to date~\cite{hadian2018end, luscher2019rwth, zhou2020rwth, li2020comparison}.
\begin{figure}[t]
\vspace{-3mm}
  \centering
  \centerline{\includegraphics[width=8cm]{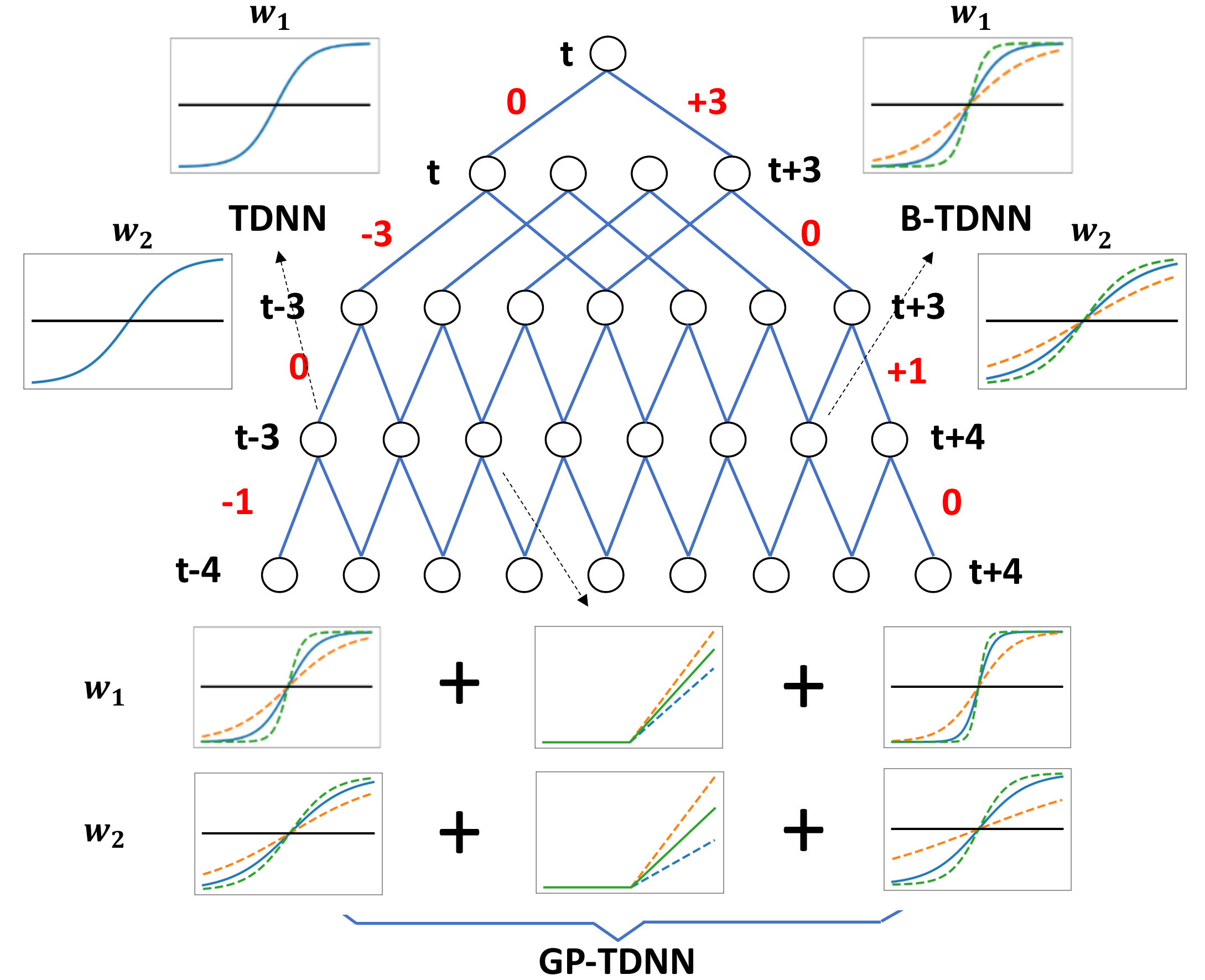}}
\caption{An example TDNN architecture with the option of using standard TDNN, Bayesian TDNN and Gaussian Process TDNN. In this example, the input dimension of all hidden nodes is assumed to be two. $w_1$ and $w_2$ are the corresponding weights for each input dimension of a hidden node. Conventional TDNN systems use fixed value, deterministic estimation of the weight parameters $w_1$, $w_2$ (Top left). B-TDNN systems of Sec.~\ref{bayesian_est_tdnn} use latent weight posterior distributions to account for model uncertainty. The GP-TDNN systems of Sec.~\ref{gp_est_tdnn} use latent weight posterior distributions for three basis activations of varying non-linearity to be combined over, thus considering uncertainty over both the weight parameters $w_1$, $w_2$ and the choice of hidden activation functions.}
\label{fig: TDNN}
\vspace{-3mm}
\end{figure}

TDNNs can be considered as a special form of one-dimensional convolutional neural networks (CNNs)~\cite{lecun1995convolutional} when parameters are tied across different time steps. An example TDNN model is shown in Fig.~\ref{fig: TDNN}. The bottom layers of TDNNs are designed to learn a narrower temporal context span, while the higher layers to learn wider, longer range temporal contexts. One important type of hyper-parameters in TDNN models controlling its temporal modelling ability is the left and right splicing context offsets. These alter the temporal context ranges effectively learned in each TDNN hidden layer. The splicing context offsets used in the example of Fig.~\ref{fig: TDNN} are \{-1,0\} \{0,1\} \{-3,0\} \{0,3\} from the bottom to the top layer. In this paper, we adopt a factored form of TDNN model structure~\cite{povey2018semi}, which compresses the weight matrix by using semi-orthogonal matrix decomposition.

\section{Bayesian Estimation of TDNN}
\label{bayesian_est}
This section presents the LF-MMI based sequence level discriminative estimation schemes for Bayesian TDNNs (B-TDNNs), Gaussian Process TDNNs (GP-TDNNs) and Variational TDNNs (V-TDNNs). In addition, dropout is re-formulated as a special case of Bayesian TDNN systems, before being further extended into a more generalized form and integrated with the full Baysian TDNN systems.
\vspace{-3mm}
\subsection{Bayesian TDNN}
\label{bayesian_est_tdnn}
For any cost error function using the cross-entropy or the sequence training criterion, for example, MMI~\cite{bahl1986maximum}, the same back-propagation algorithm in the gradient chain can be applied as in Eqn. (\ref{eq: chain rule}). The only term needs to be changed is the first part in the chain, which is modified into the specific error cost function gradient w.r.t the last layer outputs $\frac{\partial F}{\partial h_{j}^{L}}$ in different tasks. 
\begin{equation}
\label{eq: chain rule}
\setlength{\abovedisplayskip}{4pt}
\setlength{\belowdisplayskip}{4pt}
\begin{aligned}
\nabla_{\theta_{i}}^{l} F=\sum_{j}\frac{\partial F}{\partial h_{j}^{L}}\underbrace{\left\{\sum_{k}\frac{\partial h_{j}^{L}}{\partial h_{k}^{L-1}} \dots \left[\sum_{i}\frac{\partial h_{i}^{l}}{\partial \theta_{i}^{l}}\right]\right\}}_{\textbf{gradient chain}}
\end{aligned}
\end{equation}
where $\theta_{i}^{(l)}\!=\!w_{i}^{(l)}$ corresponds to the $i$-th node dependent parameter in the $l$-th layer, $h_{i}^{(l)}$ is the $i$-th hidden node output in the $l$-th layer and $F$ is the error cost function, for example, the MMI criterion~\cite{bahl1986maximum} in Eqn. (\ref{eq: MMI criterion}).
\begin{align}
\label{eq: MMI criterion}
    F_\text{MMI}(\mathbf{D};\mathbf{\Theta})\!=\!\sum_r\log\frac{p(\mathbf{O}_r|\mathbf{H}_r)^kP(\mathbf{H}_r)}{\sum_{\mathbf{H}_r^{'}}p(\mathbf{O}_r|\mathbf{H}_r^{'})^kP(\mathbf{H}_r^{'})}
\end{align}
where $P(\mathbf{H}_r^{'})$ is the language model probability for the confusable word sequence $\mathbf{H}_r^{'}$, $\mathbf{H}_r$ is the reference word sequences, $\mathbf{\Theta}$ contains the hyper-parameters of the latent distributions in Bayesian and Gaussian Process TDNNs as well as fixed parameters inside them if they are any, $k$ is the acoustic scaling factor. The sum of the denominator is taken over all possible word sequences for utterance $r$. 

The following marginalization of the training data MMI loss function in Eqn. (\ref{eq: MMI criterion}) is optimized to infer the latent weight parameter distribution in $p(\mathbf{H}_r^{*}|\mathbf{O}_r^{*},\mathbf{D})$. 
\begin{equation}
\label{eq: loss function}
\setlength{\abovedisplayskip}{4pt}
\setlength{\belowdisplayskip}{4pt}
\begin{aligned}
F = \log\int \exp\left\{F_\text{MMI}(\mathbf{D};\mathbf{\Theta})\right\}P_r(\mathbf{w})d\mathbf{w}
\end{aligned}
\end{equation}
where $\mathbf{w}\in\mathbf{\Theta}$ and $P_r(\mathbf{w})$ denotes the weight prior distribution.

When the MMI criterion in Eqn. (\ref{eq: loss function}) uses no acoustic probability scaling ($k$=1) and a sufficiently large set of confusable word sequence $\mathbf{H}_r^{'}$ for each utterance in the training data, the  evidence integral in Eqn. (\ref{eq: loss function}) is equivalent to a marginalization of the \textit{conditional maximum likelihood} of the reference word sequences ${\mathbf{H}_r}$ given ${\mathbf{O}_r}$.

The commonly used variational inference is used to approximate the integration in Eqn. (\ref{eq: loss function}). Instead of explicitly computing the posterior distribution, the evidence lower bound is first derived by Jesen's inequality in Eqn. (\ref{eq: ELBO}). Then we directly optimize the evidence lower bound to find a variational distribution $q(\mathbf{w})$ to approximate the posterior distribution. The first term in the evidence lower bound of Eqn. (\ref{eq: ELBO}) can be approximated with Monte Carlo sampling method in Eqn. (\ref{eq: sampling}). Further rearranging the second KL term in the lower bound in Eqn. (\ref{eq: ELBO}) allows the hyper-parameters $\pmb{\mu}$ and $\pmb{\sigma}$ to be differentiable and updated. 
\begin{equation}
\label{eq: ELBO}
\setlength{\abovedisplayskip}{4pt}
\setlength{\belowdisplayskip}{4pt}
\begin{aligned}
F&\ge \int q(\mathbf{w})F_\text{MMI}(\mathbf{D};\mathbf{\Theta})d\mathbf{w}-\text{KL}(q(\mathbf{w})\|P_r(\mathbf{w}))\\
&=\mathcal{L}_1^\text{MMI}-\mathcal{L}_2^\text{MMI}=\mathcal{L^\text{MMI}}
\end{aligned}
\end{equation}
where $q(\mathbf{w})$ is the variational approximation of the parameter posterior distribution $p(\mathbf{w}|\mathbf{D})$, $\text{KL}(q\|P_r)$ is the Kullback-Leibler (KL) divergence between $q$ and $P_r$. For simplicity, both $q$ and $P_r$ are assumed to be Gaussian distributions,
\begin{equation}
\label{bayesian_variational_dist}
\setlength{\abovedisplayskip}{4pt}
\setlength{\belowdisplayskip}{4pt}
\begin{aligned}
q(\mathbf{w})=\mathcal N(\pmb{\mu}, \pmb{\sigma}^2),\quad P_r(\mathbf{w})=\mathcal N(\pmb{\mu}_r, \pmb{\sigma}_r^2)
\end{aligned}
\end{equation}

The first term $\mathcal{L}_1^\text{MMI}$ in Eqn. (\ref{eq: ELBO}) can be efficiently approximated by Monte Carlo sampling method. The integrand is re-parameterized so that it does not depend on the $\pmb{\mu}$ and $\pmb{\sigma}$ directly, but instead on the standard normal distribution $\pmb{\epsilon}$.
\begin{equation}
\label{eq: sampling}
\setlength{\abovedisplayskip}{4pt}
\setlength{\belowdisplayskip}{4pt}
\begin{aligned}
\mathcal{L}_1^\text{MMI}&\approx
\frac{1}{N}\sum_{k=1}^{N}F_\text{MMI}(\mathbf{D};\mathbf{\Theta},\mathbf{w}_k)\\
\mathbf{w}_k&=\pmb{\mu}+\pmb{\sigma}\odot\pmb{\epsilon}_k
\end{aligned}
\end{equation}
where $\pmb{\epsilon}_{k}=\mathcal N(\mathbf{0}, \mathbf{I})$ is the $k$-th sample in the total $N$ samples and $\odot$ denotes the Hadamard product.

The KL divergence between $q$ and $P_r$ of the second term $\mathcal{L}_2^\text{MMI}$ can be simplified into an analytical form as follows.
\begin{equation}
\setlength{\abovedisplayskip}{4pt}
\setlength{\belowdisplayskip}{4pt}
\begin{aligned}
\mathcal{L}_2^\text{MMI}=\sum_{j}\left\{\log\frac{\sigma_{r,j}}{\sigma_{j}}+\frac{{\sigma}_{j}^2+(\mu_{j}-\mu_{r,j})^2}{2\sigma_{r,j}^2}-\frac{1}{2}\right\}
\end{aligned}
\end{equation}
where $\mu_{j}$ and $\sigma_{j}$ are the $j$-th component of variational posterior distribution hyper-parameters $\pmb{\mu}$, $\pmb{\sigma}$, $\mu_{r,j}$ and $\sigma_{r,j}$ are the $j$-th component of prior distribution hyper-parameters $\pmb{\mu}_r$ and $\pmb{\sigma}_r$ respectively. 

The gradients w.r.t the hyper-parameters $\mu_j, \sigma_j$ are given as below.
\begin{equation}
\label{eq: gradient statics}
\setlength{\abovedisplayskip}{4pt}
\setlength{\belowdisplayskip}{4pt}
\begin{aligned}
\frac{\partial{\mathcal{L^\text{MMI}}}}{\partial{\mu_j}}\!=\!\frac{1}{N}\!\sum_{k=1}^{N}\!\frac{\partial{F_\text{MMI}(\mathbf{D};\mathbf{\Theta},\pmb{\epsilon}_k)}}{\partial{\mu_j}}\!-\!\frac{\!\mu_{j}\!-\!\mu_{r,j}}{\sigma_{j}^2}\\
\frac{\partial{\mathcal{L^\text{MMI}}}}{\partial{\sigma_j}}\!=\!\frac{1}{N}\!\sum_{k=1}^{N}\!\frac{\partial{F_\text{MMI}(\mathbf{D};\mathbf{\Theta},\pmb{\epsilon}_k)}}{\partial{\sigma_j}}\!-\!\frac{\sigma_j^2\!-\!\sigma_{r,j}^2}{\sigma_j\sigma_{r,j}^2}\!
\end{aligned}
\end{equation}
where the gradients required by the right hand side of Eqn. (\ref{eq: gradient statics}) can be directly calculated using the standard back-propagation method.
\vspace{-4mm}
\subsection{Bayesian Dropout TDNN}
\label{bd_est_tdnn}
Dropout is a standard technique widely used in deep learning to avoid overfitting~\cite{srivastava2014dropout}. It can be viewed as a special form of Bayesian TDNN systems when variational distribution $q(\mathbf{w})$ is written as the following form.
\begin{equation}
\label{eq: dropout_density}
\setlength{\abovedisplayskip}{4pt}
\setlength{\belowdisplayskip}{4pt}
\begin{aligned}
q(\mathbf{w})=a\delta(\mathbf{w}) + (1-a)\mathcal N(\mathbf{0}, \pmb{\sigma}_1^2)
\end{aligned}
\end{equation}
where $a$ is the interpolation weight of two component distributions and $\pmb{\sigma}_1$ is fixed to be a small constant value, for example, $\exp(-3)$. $\delta(\mathbf{w})$ is the delta function taking the value of 1 when using the weight parameter $\mathbf{w}$. Traditionally, $a$ is parametrized as a Bernoulli random variable. In this case, $\mathbf{w}$ either keeps its original value with probability $a$, or is replaced by a dropout sample drawn from the $\mathcal N(\mathbf{0}, \pmb{\sigma}_1^2)$ with probability $1-a$.
It is clear that the first variational distribution component in the standard Dropout of Eqn.~(\ref{eq: dropout_density}) can not be Bayesian estimated. In order to generalize it and fully integrate it into the Bayesian TDNN system training process, the following Bayesian Dropout~\cite{DBLP:conf/icml/GalG16} in Eqn.~(\ref{eq: dropout_density}) can be used.
\begin{equation}
\setlength{\abovedisplayskip}{4pt}
\setlength{\belowdisplayskip}{4pt}
\label{eq: bayes_dropout_density}
\begin{aligned}
q(\mathbf{w})=a\mathcal N(\pmb{\mu}, \pmb{\sigma}_0^2) +(1-a)\mathcal N(\mathbf{0}, \pmb{\sigma}_1^2)
\end{aligned}
\end{equation}
where hyper-parameter $\pmb{\sigma}_0$ is learned by variational inference as in B-TDNN while $\pmb{\sigma}_1$ is fixed as a small constant value.

When using Monte Carlo sampling to approximate the first term of the evidence lower bound in Eqn.~(\ref{eq: ELBO}), the corresponding weight samples for Bayesian Dropout TDNNs are modified as given in Eqn.~(\ref{eq: bayes_dropout_interpolation}).
\begin{equation}
\setlength{\abovedisplayskip}{4pt}
\setlength{\belowdisplayskip}{4pt}
\label{eq: bayes_dropout_interpolation}
\begin{aligned}
\mathbf{w}_k&=a(\pmb{\mu}+\pmb{\sigma_0}\odot\pmb{\epsilon}_k) + (1-a)\pmb{\sigma_1}\odot\pmb{\epsilon}_k\\
\end{aligned}
\end{equation}
where $a$ is the interpolation weight fixed to be 0.5 on all our experiments, $\pmb{\epsilon}_k$ is the $k$-th sample drawn from $\pmb{\epsilon}_{k}=\mathcal N(\mathbf{0}, \mathbf{I})$ and $\odot$ denotes the Hadamard product. 

With the variational distribution $q(\mathbf{w})$ defined in Eqn. (\ref{eq: bayes_dropout_density}) and the prior distribution $P_r(\mathbf{w})=\mathcal N(\pmb{\mu}_r, \pmb{\sigma}_r^2)$, the second KL divergence of the second term $\mathcal{L}_2^\text{MMI}$ in Eqn.~(\ref{eq: ELBO}) can also be approximated as~\cite{DBLP:conf/icml/GalG16}:
\begin{equation}
\setlength{\abovedisplayskip}{4pt}
\setlength{\belowdisplayskip}{4pt}
\begin{aligned}
\mathcal{L}_2^\text{MMI}&\approx a\sum_{j}\left\{\frac{{\sigma}_{0,j}^2+(\mu_{j}-\mu_{r,j})^2}{2\sigma_{r,j}^2}-\log(\sigma_{0,j})\right\}\\
&+(1-a)\sum_{j}\left\{\frac{{\sigma}_{1,j}^2}{2\sigma_{r,j}^2}-\log(\sigma_{1,j})\right\}-C
\end{aligned}
\end{equation}
where $\mu_{j}$ and $\sigma_{i,j}$ are the j-th components of $\pmb{\mu}$ and $\pmb{\sigma}_{i}$, $C$ is a constant term. Note that the KL divergence term approximated in Eqn. (19) is multiplied by the probability of the weights keeping the values. If the dropout probability is set to be zero ($a=1$), it leads to the same form of Bayesian TDNNs using the variational distribution of Eqn.~(\ref{bayesian_variational_dist}).
\vspace{-3mm}
\subsection{Gaussian Process TDNN}
\label{gp_est_tdnn}
Gaussian Process time delay neural networks (GP-TDNN) model can be viewed as a specific type of the more general Gaussian Process neural networks (GPNNs) that are introduced in Sec.~\ref{gpnn}. The connection between Gaussian Processes (GP) and GP-TDNNs, lies in the fact that for each hidden node of a TDNN, a weight space view~\cite{rasmussen2006gaussian} of a Gaussian Processes expressed as a interpolation over Sigmoid, Tanh, ReLU basis activation outputs, as in Eqn. (\ref{eq: gpnn0}) of Sec.~\ref{gpnn}, is used to replace the use of a single ReLU activation function of fixed value parameters in a standard TDNN model. This corresponds to the GP-TDNN0 model shown in Tab.~\ref{uncertainty_table}. Further consideration over the modelling uncertainty associated with either the basis activation coefficients $\pmb{\lambda}$, or the basis activation internal weight parameters $\mathbf{w}$, or both of these, leads to the other GP-TDNN variants (GP-TDNN1, GP-TDNN2 and GP-TDNN3) shown in Tab.~\ref{uncertainty_table}.

Similar to the variational inference procedure used in Sec.~\ref{bayesian_est_tdnn} for Byesian TDNNs, the evidence lower bound in Eqn. (\ref{eq: gp_ELBO}) is used to approximate the MMI criterion marginalization $F$ over both $\mathbf{w}$ and $\pmb{\lambda}$.
\begin{equation}
\label{eq: gp_ELBO}
\setlength{\abovedisplayskip}{4pt}
\setlength{\belowdisplayskip}{4pt}
\begin{aligned}
F&\ge \iint q(\mathbf{w})q(\pmb{\lambda})F_\text{MMI}(\mathbf{D};\mathbf{\Theta})d\mathbf{w}d\pmb{\lambda}\\
&-\text{KL}(q(\mathbf{w})\|P_r(\mathbf{w}))-\text{KL}(q(\pmb{\lambda})\|P_r(\pmb{\lambda}))\\
&=\mathcal{L}_1^\text{MMI}-\mathcal{L}_2^\text{MMI}-\mathcal{L}_3^\text{MMI}=\mathcal{L^\text{MMI}}
\end{aligned}
\end{equation}
where $\{\pmb{\lambda}, \mathbf{w}\}\in\mathbf{\Theta}$  and we assume the statistical independence between $\mathbf{w}$ and $\pmb{\lambda}$ holds. $q(\mathbf{w})$ and $q(\pmb{\lambda})$ are the variational approximations of the parameter posterior distribution $p(\mathbf{w}|\mathbf{D})$ and basis coefficient posterior distribution $p(\pmb{\lambda}|\mathbf{D})$ respectively. $\text{KL}(q\|P_r)$ is the Kullback-Leibler (KL) divergence between $q$ and prior distribution $P_r$. Following the settings used in Byesian TDNNs of Sec.~\ref{bayesian_est_tdnn}, $q$ and $P_r$ are both set to be Gaussian distributions and the first term $\mathcal{L}_1^\text{MMI}$ is calculated by Monte Carlo sampling method. 

\begin{figure}[t]
\vspace{-3mm}
  \centering
  \centerline{\includegraphics[height=3.5cm]{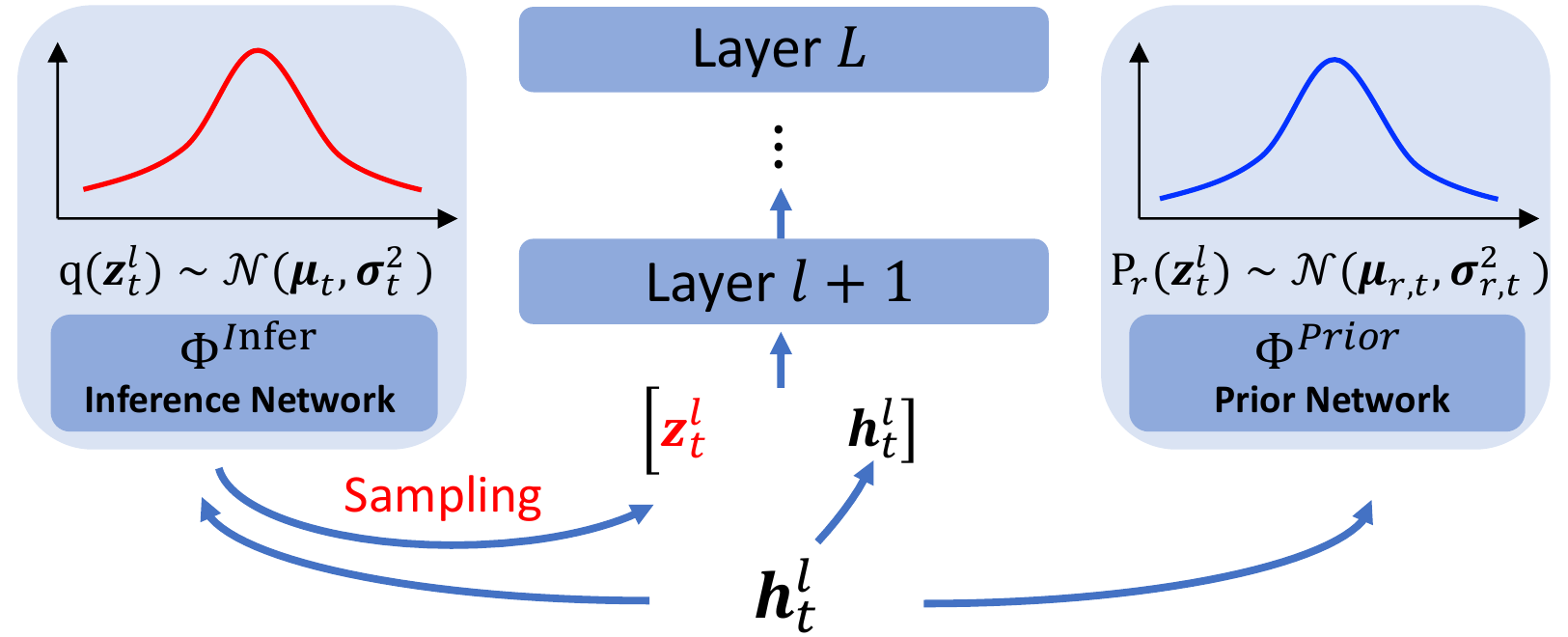}}
\caption{An example Variational TDNN. Note that the output vector of $l$-th layer $\mathbf{h}_t^l$ is used as the input of the inference network $\Phi^{Infer}$ and prior network $\Phi^{Prior}$ to calculate the mean and variance of the latent variable variational and prior distributions respectively. Then we concatenate the latent variable $\mathbf{z}_t^l$ sampled from the inference network $\Phi^{Infer}$ with $\mathbf{h}_t^l$ as the input of the ($l$+1)-th layer.}
\label{fig: VNN}
\vspace{-3mm}
\end{figure}

\vspace{-3mm}
\subsection{Variational TDNN}
\label{v_est_tdnn}
In variational TDNNs (V-TDNNs), the MMI cost function is marginalized over a sequence level hidden outputs meta-vector $\mathbf{Z}=[\mathbf{z}_1, \mathbf{z}_2, \cdots, \mathbf{z}_T]$, where the time instance level random hidden output vectors  $[\mathbf{z}_1, \mathbf{z}_2, \cdots, \mathbf{z}_T]$ are assumed to be independent among themselves. This leads to the following marginalized cost function
\begin{equation}
\label{eq: variational loss function}
\setlength{\abovedisplayskip}{4pt}
\setlength{\belowdisplayskip}{4pt}
\begin{aligned}
F &= \log\int \exp\left\{F_\text{MMI}(\mathbf{D};\mathbf{\Theta},\mathbf{Z})\right\}P_r(\mathbf{Z})d\mathbf{Z}\\
&= \sum_{t=1}^T \log\int \exp\left\{F_\text{MMI}(\mathbf{D};\mathbf{\Theta},\mathbf{z}_t)\right\}P_r(\mathbf{z}_t)d\mathbf{z}_t
\end{aligned}
\end{equation}
where $\mathbf{z}_t$ is the latent variable at time $t$ and $P_r(\mathbf{z}_t)$ denotes the prior distribution of the latent variable.

Then, the variational lower bound is derived to approximate the marginalization of MMI criterion $F$ in Eqn. (\ref{eq: vtdnn_variational_ELBO}).
\begin{equation}
\label{eq: vtdnn_variational_ELBO}
\setlength{\abovedisplayskip}{4pt}
\setlength{\belowdisplayskip}{4pt}
\begin{aligned}
F&\ge \sum_{t}\int q(\mathbf{z}_t)F_\text{MMI}(\mathbf{D};\mathbf{\Theta})d\mathbf{z}_t-\sum_{t}\text{KL}(q(\mathbf{z}_t)\|P_r(\mathbf{z}_t))\\
&=\mathcal{L}_1^\text{MMI}-\mathcal{L}_2^\text{MMI}=\mathcal{L^\text{MMI}}
\end{aligned}
\end{equation}
where $q(\mathbf{z}_t)$ is the variational approximation of the posterior distribution $p(\mathbf{z}_t|\mathbf{D})$ and $P_r(\mathbf{z}_t)$ is the prior distribution. As shown in Fig.~\ref{fig: VNN}, 
\begin{equation}
\setlength{\abovedisplayskip}{4pt}
\setlength{\belowdisplayskip}{4pt}
\begin{aligned}
q(\mathbf{z}_t)=\mathcal N(\pmb{\mu}_t,\pmb{\sigma}_t^2),\quad P_r(\mathbf{z}_t)=\mathcal N(\pmb{\mu}_{r,t}, \pmb{\sigma}_{r,t}^2)
\end{aligned}
\end{equation}
where hyper-parameters $\pmb{\mu}_t$, $\pmb{\sigma}_t^2$ are calculated from an inference network $\Phi^{Infer}(\mathbf{h}_t)$, hyper-parameters $\pmb{\mu}_{r,t}$, $\pmb{\sigma}_{r,t}^2$ are computed by a prior network $=\Phi^{Prior}(\mathbf{h}_t)$.

In common with the estimation procedures used in B-TDNN, $\mathcal{L^\text{MMI}}$ is further approximated by Monte Carlo sampling, i.e.,
\begin{equation}
\label{eq: vtdnn_L1_monte_carlo}
\setlength{\abovedisplayskip}{4pt}
\setlength{\belowdisplayskip}{4pt}
\begin{aligned}
\mathcal{L^\text{MMI}}&=\mathcal{L}_1^\text{MMI}-\mathcal{L}_2^\text{MMI}\\
&\approx\sum_{t}\frac{1}{N}\sum_{k=1}^{N} F_\text{MMI}(\mathbf{D};\mathbf{\Theta},\pmb{\mu}_t+\pmb{\sigma}_t\odot\pmb{\epsilon}_{k})\\
&-\sum_{t}\text{KL}(q(\mathbf{z}_t)\|P_r(\mathbf{z}_t))
\end{aligned}
\end{equation}
where $\mathbf{z}_t=\pmb{\mu}_t+\pmb{\sigma}_t\odot\pmb{\epsilon}_{k}$, $\pmb{\epsilon}_{k}=\mathcal N(\mathbf{0}, \mathbf{I})$ is the $k$-th sample in the total $N$ samples and $\odot$ denotes the Hadamard product.

\vspace{-2mm}
\subsection{Implementation Details}
\label{im_detail}
The performance and efficiency of the proposed Bayesian, Gaussian Process and Variational TDNN systems in Sec.~\ref{bayesian_est_tdnn} to \ref{v_est_tdnn} are affected by the following set of implementation details. Striking a sensible balance between performance and computational cost is crucial for practical implementation of these systems. 

1) {\bf Choice of prior distribution:} 
When training Bayesian learning based models, a suitable choice of parameter prior needs to be set. In our experiments, we set the priors for various Bayesian learned LF-MMI TDNN systems to be based on the comparable converged standard fixed-parameter TDNN systems. In addition, all the other parameters in the Bayesian learning based LF-MMI TDNN models are initialized using the parameters obtained from the comparable half-trained standard TDNN systems. The combination of these two settings in practice was found to yield a good balance between convergence speed and performance.

2) {\bf Modelling uncertainty at different layers}
Applying Bayesian estimation of all layers inside TDNN systems is highly expensive in both model training and evaluation. It is well known that deep neural networks including TDNNs are powerful models that are capable to produce denoised and invariant features in their higher layers for accurate classification of the inputs. It is therefore expected that the modelling uncertainty associated with the lower layers of TDNN systems will be much larger than those of the higher layers. This is confirmed in our experiments that are to be presented in Sec.~\ref{exp_swbd75}. In practice, it is found that Bayesian estimation only needs to be applied to the first TDNN hidden layer where the largest modelling uncertainty is expected, while further applying Bayesian estimation of any subsequent higher layers produces no further performance improvement. 

3) {\bf Parameter tying for variational distributions}
The extensive use of variational distributions during Bayesian inference in Sec.~\ref{bayesian_est_tdnn} to \ref{v_est_tdnn} leads to a large number of latent distribution hyper-parameters to be estimated and stored. In order to ensure the number of free parameters in the proposed Bayesian and Gaussian Process TDNN systems to be comparable to that of the standard TDNN systems, the variational distribution variance $\pmb{\sigma}$ is shared among all the hidden nodes of the same layer for Bayesian and Gaussian Process TDNN systems. In addition, we further share the latent distribution over the weight parameters $\mathbf{w}$ across all basis activation functions of Eqn. (\ref{eq: gp_general}) in GP-TDNN systems to control the overall system complexity.

4) {\bf Parameter sampling in inference}
The inference algorithms of all the Bayesian estimated TDNN systems presented in Sec.~\ref{bayesian_est_tdnn} to \ref{v_est_tdnn} require the use of Monte Carlo sampling to approximate the respective parts of their lower bounds computing the MMI criterion expectation in  Eqn. (\ref{eq: sampling}) and Eqn. (\ref{eq: vtdnn_L1_monte_carlo}) given the variational distributions. The resulting inference cost during model training is therefore linearly increased with respect to the number of samples being drawn as in Tab.~\ref{uncertainty_table}. Experimental results in Tab.~\ref{uncertainty_sample} further show that only a marginal difference in Word Error Rate (WER) was observed by drawing more samples (two and three samples) in the forward pass of Bayesian TDNN systems. In order to maintain the Bayesian learned TDNN systems’ overall computational cost during model training comparable to that of the conventional TDNNs as shown in line 2 and 5-9 in Tab.~\ref{uncertainty_table}, only one sample is drawn in Eqn. (\ref{eq: sampling}) and Eqn. (\ref{eq: vtdnn_L1_monte_carlo}) for all the Bayesian estimated TDNN systems presented in this paper. The KL term in Eqn. (\ref{eq: ELBO}), Eqn. (\ref{eq: gp_ELBO}) and Eqn. (\ref{eq: vtdnn_variational_ELBO}) is set to be proportional to the batch size. During evaluation, the inference of Bayesian, Gaussian Process and Variational TDNNs in Eqn. (\ref{eq: bayesian_likelihood}), Eqn. (\ref{eq: gp_likelihood}) and Eqn. (\ref{eq: variational_likelihood}) are efficiently approximated by computing the expectation of the model parameters or the latent variables using the respective posterior distributions. For example, during recognition time, the weight parameters $\mathbf{w}$ in the B-TDNN systems are approximated by the mean of their latent distribution given as follows:
\begin{equation}
\begin{aligned}
&\int p(\mathbf{H}_r^{*}|\mathbf{O}_r^{*},\mathbf{w}) p({\mathbf{w}}|\mathbf{D})d\mathbf{w}\approx p\left(\mathbf{H}_r^{*}|\mathbf{O}_r^{*},\mathbf{E}\left[\mathbf{w}|\mathbf{D}\right]\right)
\end{aligned}
\end{equation}
Thus, the speed ratio relative to the standard TDNN system in Tab.~\ref{uncertainty_table} is approximately 1.0 at the test stage.

5) {\bf System description}
Following the above implementation details, the description of a set of Bayesian, Gaussian Process and Variational TDNN systems in terms of their respective forms of uncertainty modelling, number of free parameters after tying and the speed ratio relative to the standard TDNN systems in training and evaluation is presented in Tab.~\ref{uncertainty_table}. In the table, four variants GP-TDNN systems by considering no uncertainty (GP-TDNN0), or the uncertainty associated with either the activation basis coefficients $\pmb{\lambda}$ (GP-TDNN1), or the activation internal weight parameters $\mathbf{w}$ alone (GP-TDNN2), or the uncertainty associated with both $\mathbf{w}$ and $\pmb{\lambda}$ (GP-TDNN3), are also shown.

As shown in Tab.~\ref{uncertainty_table}, by assuming the input vector size as $a$, the number of hidden nodes as $b$ and the latent variable $\mathbf{z}$ dimension as $c$ for a single layer, each standard TDNN layer has a total of $ab$ parameters in the weight matrix $\mathbf{w}$. With the aforementioned parameter tying, each B-TDNN layer has a total of $ab$ latent distribution parameters for the mean $\pmb{\mu}$, and the number of parameters in the shared variance being $a$. Compared with a standard TDNN layer, a GP-TDNN0 layer has a total of $ab$ parameters for the weight matrix, plus $3b$ additional parameters for the basis activation coefficients $\pmb{\lambda}$. Compared with a GP-TDNN0 layer, a GP-TDNN1 layer requires 3 more parameters for the shared variance term of the latent distribution over $\pmb{\lambda}$. By sharing the latent distribution over $\mathbf{w}$ across all basis activation functions of Eqn. (\ref{eq: gp_general}), a GP-TDNN2 layer only needs a total of $a$ more parameters for the shared variance term of the distribution over $\mathbf{w}$. If we model the uncertainty over both the basis activation coefficients $\pmb{\lambda}$ and weight parameters $\mathbf{w}$ inside the activation functions, the GP-TDNN3 layer has a total of $3+a$ more parameters than a GP-TDNN0 layer. For the Variational TDNN layer modelling the hidden output uncertainty as shown in Fig.~\ref{fig: VNN}, it requires $4ac$ parameters for the inference network $\Phi^{Infer}$ and prior network $\Phi^{Prior}$ to generate the mean and variance of the latent vatiables $\mathbf{z}$, plus additional $ab+cb$ parameters for the weight matrix $\mathbf{w}$.

In addition, when using the above efficient sampling during inference for model training (as low as one sample drawn) and evaluation, the Bayesian, Gaussian Process and Variational TDNN systems only require a moderate increase in system training time of 10\%-30\% over the standard TDNN baseline systems during training, while their computational complexity is comparable to that of standard TDNN systems during the testing stage. 

\begin{table}[t]
\vspace{-3mm}
\caption{Description of Bayesian, Gaussian Process and Variational TDNN layer in terms of their respective forms of number of samples, uncertainty modelling, number of free parameters after tying and the speed ratio relative to the standard TDNN layer in training and evaluation, by assuming the the input vector size as $a$, the number of hidden nodes as $b$ and the latent variable $\mathbf{z}$ dimension as $c$ for a single layer. The Bayesian and Gaussian Process TDNN layers are constructed following the implementation details of Section IV-E.  }\label{uncertainty_table}
\centering
\setlength{\tabcolsep}{1.8mm}{
\begin{tabular}{c|c|ccc|ccc|c|cc}
\hline \hline
\multirow{2}*{Layer} & \#Sa- &\multicolumn{3}{c|}{Uncertainty} & \multicolumn{3}{c|}{\#Param} & \multicolumn{2}{c}{Speed Ratio}\\
\cline{3-10}
~ & mple & $\pmb{\lambda}$ & $\mathbf{w}$ & $\mathbf{z}$ & $\pmb{\lambda}$ & $\mathbf{w}$ & $\mathbf{z}$ & Train & Test\\
\hline \hline 
TDNN & 1 & \xmark & \xmark & \xmark & 0 & ab & 0 & 1.0 & 1.0\\
\hline \hline 
B-TDNN & 1 & \xmark & \cmark & \xmark & 0 & ab+a & 0 & 1.2 & 1.0\\
B-TDNN & 2 & \xmark & \cmark & \xmark & 0 & ab+a & 0 & 1.4  & 1.0\\
B-TDNN & 3 & \xmark & \cmark & \xmark & 0 & ab+a & 0 & 1.6  & 1.0\\
\hline \hline 
GP-TDNN0 & 1 & \xmark & \xmark & \xmark & 3b & ab & 0 & 1.1 & 1.0\\
GP-TDNN1 & 1 & \cmark & \xmark & \xmark & 3b+3 & ab & 0 & 1.1 & 1.0\\
GP-TDNN2 & 1 & \xmark & \cmark & \xmark & 3b & ab+a & 0 & 1.2 & 1.0\\
GP-TDNN3 & 1 & \cmark & \cmark & \xmark & 3b+3 & ab+a & 0 & 1.2 & 1.0\\
\hline \hline 
V-TDNN & 1 & \xmark & \xmark & \cmark & 0 & ab+cb & 4ac & 1.3 & 1.0\\
\hline \hline 
\end{tabular}}
\vspace{-3mm}
\end{table}

\begin{table}[h]
\vspace{-3mm}
\caption{Performance (WER\%) comparison between TDNN, B-TDNN systems constructed using the 75-hour Switchboard training subset by drawing varying number of samples (1,2 and 3). The WERs were evaluated on the \textbf{HUB5' 00}, \textbf{RT03S} and \textbf{RT02} test sets. $^\dagger$ denotes a statistically significant difference is obtained over the TDNN baseline system (line 1). (SWB1 and CHM denote the Switchboard and Callhm subsets of the Hub5' 00 test set; FSH and SWB2 denote the Fisher and Switchboard subsets of the RT03S test set; SWB3, SWB4 and SWB5 denote three Switchboard subsets in the Rt02 test set.)}
\label{uncertainty_sample}
\centering
\setlength{\tabcolsep}{0.7mm}{
\begin{tabular}{c|c|cc|cc|ccc|c}
\hline\hline
\multirow{2}*{System} & \#Sam- & \multicolumn{2}{c}{Hub5' 00} & \multicolumn{2}{c}{Rt03S} & \multicolumn{3}{c|}{Rt02} & Total\\
\cline{3-9}
~ & ple & SWB1 & CHM & FSH  & SWB2 & SWB3 & SWB4 & SWB5 & Avg\\
\hline\hline
TDNN & - & 12.2 & 24.2 & 16.6 & 26.3 & 14.5 & 19.8 & 27.6 & 20.7\\
\hline\hline
\multirow{3}*{B-TDNN} & 1 & $11.7^\dagger$ & $\mathbf{23.3}^\dagger$ & $\mathbf{15.6}^\dagger$ & $\mathbf{25.0}^\dagger$ & 14.3 & \textbf{19.2}$^\dagger$ & 25.8$^\dagger$ & 19.7$^\dagger$\\
 & 2 & 12.0 & 23.6$^\dagger$ & 15.7$^\dagger$ & 25.4$^\dagger$ & 14.5 & 19.3$^\dagger$ & 25.9$^\dagger$ & 19.9$^\dagger$\\
 & 3 & \textbf{11.6}$^\dagger$ & 23.7$^\dagger$ & 15.9$^\dagger$ & \textbf{25.0}$^\dagger$ & 14.3 & \textbf{19.2}$^\dagger$ & \textbf{25.7}$^\dagger$ & 19.8$^\dagger$\\
\hline\hline
\end{tabular}}
\vspace{-3mm}
\end{table}

\section{Experiments}

This section is organized as follows. Firstly, in Section V-A, the performance of various Bayesian, Gaussian Process and Variational LF-MMI TDNN systems constructed using a 75-hour subset of the LDC Switchboard I data are evaluated. This initial set of experiments serve to confirm the implementation details and settings given in Sec.~\ref{im_detail} suitable to use for the subsequent larger experiments in the rest of this paper. Secondly, in Sec.\ref{exp_swbd900}, the main set of experiments are conducted on a full 900-hour speed-perturbed Switchboard corpus to fully evaluate the performance of the Bayesian estimated LF-MMI TDNN systems proposed in Sec.~\ref{bayesian_est}. Performance comparison against the baseline LF-MMI TDNN system which used multiple regularization methods (F-smoothing, L2 norm penalty, natural gradient, model averaging and dropout), in addition to i-Vector plus learning hidden unit contribution (LHUC) based speaker adaptation and Kaldi recipe LSTM recurrent neural network language model (RNNLM) rescoring is drawn. Thirdly, in Sec.\ref{exp_hkust450}, a comparable set of experiments are conducted in a 450-hour speed-perturbed HKUST conversational Mandarin telephone speech recognition task. Finally, the performance of Bayesian estimated LF-MMI TDNN systems are further evaluated on a cross domain adaptation task which requires porting a 1000 hour LibriSpeech data trained LF-MMI TDNN system to a small DementiaBank elderly speech corpus. In all our experiments, we follow the Kaldi chain model setup\footnote{All of this is in published Kaldi code at https://github.com/kaldi-asr/kaldi/tree/master/egs/*/*/local/chain/tuning/run \_tdnn\_7q.sh}, except that we used 40-dimension filterbank features as the input features instead of the 40-dimension high-resolution Mel-frequency cepstral coefficients (MFCCs). All of our models were trained with one thread on a single NVIDIA Tesla V100 Volta GPU card. For all results presented in this paper, matched pairs sentence-segment word error (MAPSSWE) based statistical significance test was performed at a significance level $\alpha=0.05$.

\begin{figure*}[h]
\vspace{-3mm}
    \centering
    \begin{minipage}[h]{0.45\textwidth}
        \centering
        \includegraphics[width=3in]{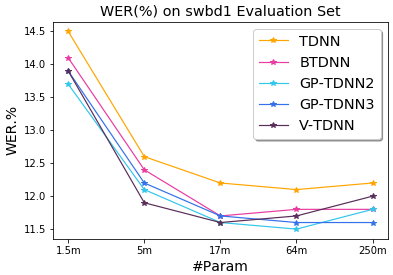}
        {\par \footnotesize a\par}
    \end{minipage}
    \hfill
    \begin{minipage}[h]{0.45\textwidth}
        \centering
        \includegraphics[width=3in]{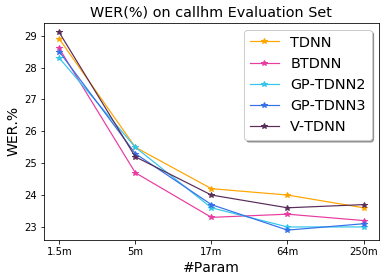}
        {\par \footnotesize b\par}
    \end{minipage}
    \hfill
    \begin{minipage}[h]{0.45\textwidth}
        \centering
        \includegraphics[width=3in]{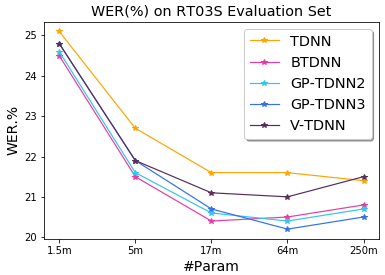}
        {\par \footnotesize c\par}
    \end{minipage}
    \hfill
    \begin{minipage}[h]{0.45\textwidth}
        \centering
        \includegraphics[width=3in]{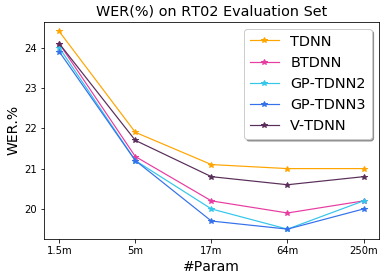}
        {\par \footnotesize d\par}
    \end{minipage}
    \caption{Performance comparison of different TDNN, B-TDNN, GP-TDNN2, GP-TDNN3 and V-TDNN systems with equal model complexity on the \textbf{HUB5' 00} (a,b), \textbf{Rt03S} (c) and \textbf{Rt02} (d) test sets. The model size (measured in the number of parameters) is increased from 1.5m to 5m, 17m, 64m and up to 250m by varying the linear, projection and affine layers dimensionality. The standard TDNN system (line 1 in Tab.~\ref{swbd75_result}) containing 17m parameters is shown in the middle of the four figures.}
    \label{fig:model_size_performace_75h}
    \vspace{-3mm}
\end{figure*}

\vspace{-3mm}
\subsection{Experiments on 75-Hour Switchboard Task}
\label{exp_swbd75}
In this part, an investigation of different full Bayesian TDNN learning variants is conducted on the 75-hour Switchboard task to verify the feasibility of implementation details and settings in Sec.~\ref{im_detail} for further experiments in the rest of this paper. We first investigate the suitable number of layers to apply Bayesian estimation. After determining which layer(s) to incorporate Bayesian modelling, we compare the performance of the Bayesian TDNN system, Bayesian Dropout TDNN system, Gaussian Process TDNN system and Variational TDNN system described in Sec.~\ref{bayesian_est_tdnn} to \ref{v_est_tdnn}. Finally, we demonstrate the robustness of different Bayesian learning based TDNN systems by varying the model sizes (by varying hidden layer dimensionality).

\begin{table}[h]
\vspace{-3mm}
\caption{Performance (WER\%) comparison of TDNN, B-TDNN, BD-TDNN, GP-TDNN and V-TDNN systems considering the uncertainty at different layers constructed using the 75-hour Switchboard training subset. The WERs were evaluated on the \textbf{HUB5' 00}, \textbf{RT03S} and \textbf{RT02} test sets. $^\dagger$ denotes a statistically significant difference is obtained over the TDNN baseline system (line 1). (SWB1 and CHM denote the Switchboard and Callhm subsets of the Hub5' 00 test set; FSH and SWB2 denote the Fisher and Switchboard subsets of the RT03S test set; SWB3, SWB4 and SWB5 denote three Switchboard subsets in the Rt02 test set.)}
\label{swbd75_result}
\centering
\setlength{\tabcolsep}{0.8mm}{
\begin{tabular}{c|c|cc|cc|ccc}
\hline\hline
\multirow{2}*{System} & \multirow{2}*{Layer} & \multicolumn{2}{c}{Hub5' 00} & \multicolumn{2}{c}{Rt03S} & \multicolumn{3}{c}{Rt02}\\
\cline{3-9}
~ & ~ & SWB1 & CHM & FSH  & SWB2 & SWB3 & SWB4 & SWB5\\
\hline\hline
TDNN & 1-15 & 12.2 & 24.2 & 16.6 & 26.3 & 14.5 & 19.8 & 27.6\\
\hline\hline
\multirow{5}*{B-TDNN} & 1 & $\mathbf{11.7}^\dagger$ & $\mathbf{23.3}^\dagger$ & $\mathbf{15.6}^\dagger$ & $\mathbf{25.0}^\dagger$ & $\mathbf{14.3}$ & 19.2$^\dagger$ & $\mathbf{25.8}^\dagger$\\
~ & 1-2 & 11.9 & 23.5$^\dagger$ & 16.0$^\dagger$ & 26.1$^\dagger$ & 14.4 & $\mathbf{18.9}^\dagger$ & 25.9$^\dagger$ \\
~ & 1-5 & 12.1 & 23.5$^\dagger$ & 16.1$^\dagger$ & 25.3$^\dagger$ & 14.4 & 19.4 & 26.0$^\dagger$ \\
~ & 1-8 & 11.9 & 23.6$^\dagger$ & 15.8$^\dagger$ & 25.4$^\dagger$ & 14.5 & 19.3 & 26.3$^\dagger$ \\
~ & 1-15 & 12.1 & 24.3 & 16.4 & 26.1 & 15.2 & 19.5 & 27.2 \\
\hline\hline
BD-TDNN & 1 & 11.8$^\dagger$ & 23.6$^\dagger$ & 15.6$^\dagger$ & 25.0$^\dagger$ & 14.4 & 19.1$^\dagger$ & 25.6$^\dagger$\\
\hline\hline
GP-TDNN0 & 1 & 11.8$^\dagger$ & 24.2 & 16.5 & 25.8$^\dagger$ & 14.5 & 19.6 & 26.7$^\dagger$ \\
GP-TDNN1 & 1 & $\mathbf{11.6}^\dagger$ & $\mathbf{23.5}^\dagger$ & 16.2$^\dagger$ & 25.4$^\dagger$ & 14.4 & 18.8$^\dagger$ & 26.6$^\dagger$ \\
GP-TDNN2 & 1 & $\mathbf{11.6}^\dagger$ & 23.6$^\dagger$ & $\mathbf{15.9}^\dagger$ & $\mathbf{25.0}^\dagger$ & $\mathbf{13.9}^\dagger$ & 18.8$^\dagger$ & 26.1$^\dagger$ \\
GP-TDNN3 & 1 & 11.7$^\dagger$ & 23.7$^\dagger$ & 16.0$^\dagger$ & 25.1$^\dagger$ & 14.1 & $\mathbf{18.4}^\dagger$ & $\mathbf{25.4}^\dagger$ \\
\hline\hline
V-TDNN & 1 & 11.6$^\dagger$ & 24.0 & 16.3 & 25.5$^\dagger$ & 14.2 & 19.6 & 27.1 \\
\hline\hline
\end{tabular}}
\vspace{-3mm}
\end{table}

{\bf Task Description:} Our 75-hour Switchboard I data consists of randomly selected 1082 conversational sides out of the 4870 speakers from the 300-hour Switchboard I corpus released by LDC (LDC97S62). On top of the Linear Discriminant Analysis (LDA) transformed Perceptual Linear Prediction (PLP) coefficients up to the second order, our baseline GMM-HMM system with 2904 tied tri-phone states was trained using Maximum Likelihood Linear Transform (MLLT)~\cite{leggetter1995maximum, gales1998maximum}. The speaker adaptive training (SAT)~\cite{anastasakos1996compact, anastasakos1997speaker, povey2008fast} approach was also applied to further generate the alignments for neural network training and the numerator lattices for LF-MMI training. For performance evaluation, a four-gram language model (LM) trained on the Switchboard and Fisher transcripts (LDC2004T19, LDC2005T19) was used to evaluate NIST HUB5'00 (LDC2002S09, LDC2002T43), RT03 (LDC2007S10) and RT02 (LDC2004S11) test sets. The performance of the LF-MMI trained standard TDNN system\footnote{All of this is in published Kaldi code at https://github.com/kaldi-asr/kaldi/tree/master/egs/swbd/s5c/local/chain/tuning/run \_tdnn\_7q.sh} is shown in line 1 of Tab.~\ref{swbd75_result}. At this stage, i-Vector \cite{dehak2011front} and speed perturbation were not incorporated.



{\bf Experimental Results and Analysis} As shown by most results in Tab.~\ref{swbd75_result}, the proposed Bayesian estimated TDNN systems except the Variational TDNN system significantly outperform the TDNN baseline system (line 1 in Tab.~\ref{swbd75_result}) across all three test sets. Several trends are listed as follows.

\begin{table*}[h]
\vspace{-3mm}
\caption{Performance (WER\%) comparison of TDNN, B-TDNN and GP-TDNN systems on the \textbf{HUB5' 00}, \textbf{RT03S} and \textbf{RT02} test sets before and after applying LHUC and RNNLM rescoring. $^\dagger$ denotes a statistically significant difference is obtained over the TDNN baseline system (line 1, 8, 14, 21). (SWB1 and CHM denote the Switchboard and Callhm subsets of the Hub5' 00 test set; FSH and SWB2 denote the Fisher and Switchboard subsets of the RT03S test set; SWB3, SWB4 and SWB5 denote three Switchboard subsets in the Rt02 test set.)}
\label{swbd900_result}
\centering
\begin{tabular}{c|c|c|c|c|c|cc|cc|ccc}
\hline\hline
 &
\multirow{2}*{System} &
\multirow{2}*{I-Vector} & \multirow{2}*{Speed perturb} &
\multirow{2}*{LHUC} & \multirow{2}*{LM} & \multicolumn{2}{c}{Hub5' 00} & \multicolumn{2}{c}{Rt03S} & \multicolumn{3}{c}{Rt02}\\
\cline{7-13}
~ & ~ & ~ & ~ & ~ & ~ & SWB1 & CHM & FSH  & SWB2 & SWB3 & SWB4 & SWB5\\
\hline\hline
1 & TDNN & \cmark & \xmark & \cmark & +RNNLM & 7.8 & 15.2 & 9.9 & 16.2 & 9.1 & 12.1 & 17.0\\
\hline
2 & B-TDNN & &  &  & & 7.6 & 15.0 & 9.8 & 15.8$^\dagger$ & 9.0 & 12.1 & \textbf{15.6}$^\dagger$ \\
3 & GP-TDNN0 &  &  &  &  & 7.5$^\dagger$ & 14.9$^\dagger$ & \textbf{9.5}$^\dagger$ & 15.7$^\dagger$ & 9.0 & 12.0 & 15.9$^\dagger$\\
4 & GP-TDNN1 & \cmark & \xmark & \cmark & +RNNLM & \textbf{7.3}$^\dagger$ & 15.0 & 9.6$^\dagger$ & \textbf{15.3}$^\dagger$ & \textbf{8.7}$^\dagger$ & 12.2 & \textbf{15.6}$^\dagger$\\
5 & GP-TDNN2 &  &  &  &  & \textbf{7.3}$^\dagger$ & \textbf{14.5}$^\dagger$ & 9.8 & 15.5$^\dagger$ & 9.0 & 12.0 & \textbf{15.6}$^\dagger$\\
6 & GP-TDNN3 &  &  &  &  & 7.5$^\dagger$ & 14.8$^\dagger$ & 9.8 & 15.6$^\dagger$ & 9.2 & 12.1 & 16.2$^\dagger$\\
7 & V-TDNN &  &  &  &  & 7.5$^\dagger$ & 14.9$^\dagger$ & 9.8 & 15.8$^\dagger$ & \textbf{8.7}$^\dagger$ & 12.3 & 16.5$^\dagger$\\
\hline\hline
8 & TDNN & \cmark & \cmark & \xmark & 4-gram & 9.7 & 18.0 & 12.6 & 19.5 & 11.5 & 15.3 & 20.0\\
\hline
9 & B-TDNN &  &  &  &  & 9.4 & 17.3$^\dagger$ & 12.1$^\dagger$ & 19.2 & 11.4 & 14.7$^\dagger$ & 19.1$^\dagger$\\
10 & GP-TDNN0 &  &  &  &  & \textbf{9.2}$^\dagger$ & 17.6$^\dagger$ & 12.0$^\dagger$ & 19.1$^\dagger$ & \textbf{10.8}$^\dagger$ & 14.6$^\dagger$ & 19.3$^\dagger$\\
11 & GP-TDNN1 & \cmark & \cmark & \xmark & 4-gram & \textbf{9.2}$^\dagger$ & \textbf{17.2}$^\dagger$ & 11.9$^\dagger$ & \textbf{19.0}$^\dagger$ & 11.0$^\dagger$ & 14.6$^\dagger$ & 19.5\\
12 & GP-TDNN2 &  &  &  &  & 9.3$^\dagger$ & \textbf{17.2}$^\dagger$ & \textbf{11.8}$^\dagger$ & 19.1$^\dagger$ & 11.0$^\dagger$ & \textbf{14.4}$^\dagger$ & \textbf{19.1}$^\dagger$\\
13 & GP-TDNN3 &  & &  &  & 9.5 & 17.3$^\dagger$ & 12.0$^\dagger$ & 19.1$^\dagger$ & 11.1$^\dagger$ & 14.8$^\dagger$ & 19.2$^\dagger$\\
14 & V-TDNN & &  &  &  & 9.5 & 17.9 & 12.5 & 19.6 & 11.5 & 15.2 & 19.6\\
\hline \hline
15 & TDNN & \cmark & \cmark & \cmark & 4-gram & 9.5 & 17.6 & 12.1 & 19.0 & 11.0 & 14.8 & 19.1\\
\hline
16 & B-TDNN &  &  &  &  & \textbf{9.2}$^\dagger$ & 17.1$^\dagger$ & 11.6$^\dagger$ & \textbf{18.1}$^\dagger$ & 10.8 & 14.0$^\dagger$ & 17.9$^\dagger$\\
17 & GP-TDNN0 &  &  &  &  & 9.0$^\dagger$ & 17.1$^\dagger$ & 11.6$^\dagger$ & \textbf{18.1}$^\dagger$ & \textbf{10.4}$^\dagger$ & 14.2$^\dagger$ & 17.9$^\dagger$\\
18 & GP-TDNN1 & \cmark & \cmark & \cmark & 4-gram & 9.1$^\dagger$ & 17.0$^\dagger$ & 11.9 & 18.7 & 10.7 & 14.4 & 17.9$^\dagger$\\
19 & GP-TDNN2 &  &  &  &  & 9.3 & \textbf{16.8}$^\dagger$ & 11.5$^\dagger$ & \textbf{18.1}$^\dagger$ & 10.8 & \textbf{13.9}$^\dagger$ & 18.0$^\dagger$\\
20 & GP-TDNN3 &  &  &  &  & \textbf{9.2}$^\dagger$ & 16.9$^\dagger$ & \textbf{11.3}$^\dagger$ & \textbf{18.1}$^\dagger$ & 10.6$^\dagger$ & 14.1$^\dagger$ & \textbf{17.7}$^\dagger$\\
21 & V-TDNN &  &  &  &  & 9.3 & 17.6 & 12.0 & 19.1 & 11.1 & 14.6 & 18.8\\
\hline \hline
22 & TDNN & \cmark & \cmark & \xmark & +RNNLM &  8.1 & 15.6 & 10.4 & 17.2 & 9.9 & 13.0 & 17.3\\
\hline
23 & B-TDNN &  &  &  &  & \textbf{7.7}$^\dagger$ & \textbf{14.7}$^\dagger$ & 10.2 & 16.6$^\dagger$ & 9.5 & 12.6$^\dagger$ & 16.7$^\dagger$\\
24 & GP-TDNN0 &  &  &  &  & 7.8$^\dagger$ & 15.3$^\dagger$ & 10.1$^\dagger$ & 16.6$^\dagger$ & \textbf{9.0}$^\dagger$ & 12.4$^\dagger$ & 16.8$^\dagger$\\
25 & GP-TDNN1 & \cmark & \cmark & \xmark & +RNNLM & \textbf{7.7}$^\dagger$ & 15.1$^\dagger$ & \textbf{10.0}$^\dagger$ & 16.6$^\dagger$ & 9.3$^\dagger$ & 12.5$^\dagger$ & 16.7$^\dagger$\\
26 & GP-TDNN2 &  &  &  &  & 7.9 & \textbf{14.7}$^\dagger$ & 10.1$^\dagger$ & \textbf{16.3}$^\dagger$ & 9.2$^\dagger$ & \textbf{12.3}$^\dagger$ & \textbf{16.2}$^\dagger$\\
27 & GP-TDNN3 &  &  &  &  & 7.8$^\dagger$ & 15.1$^\dagger$ & 10.1$^\dagger$ & 16.4$^\dagger$ & 9.4$^\dagger$ & \textbf{12.3}$^\dagger$ & 16.4$^\dagger$\\
28 & V-TDNN &  &  &  &  & 8.1 & 15.6 & 10.5 & 17.1 & 9.6 & 13.0 & 16.9\\
\hline \hline
29 & TDNN & \cmark & \cmark & \cmark & +RNNLM & 7.9 & 15.2 & 10.1 & 16.3 & 9.5 & 12.4 & 16.1\\
\hline
30 & B-TDNN &  &  &  &  & \textbf{7.4}$^\dagger$ & 14.6$^\dagger$ & 9.6$^\dagger$ & 15.4$^\dagger$ & 8.8$^\dagger$ & 11.9$^\dagger$ & 15.4$^\dagger$\\
31 & GP-TDNN0 &  &  &  &  & 7.5$^\dagger$ & 14.8$^\dagger$ & 9.9$^\dagger$ & 15.6$^\dagger$ & 8.9$^\dagger$ & 12.1$^\dagger$ & 15.1$^\dagger$\\
32 & GP-TDNN1 & \cmark & \cmark & \cmark &+RNNLM & 7.5$^\dagger$ & 14.9$^\dagger$ & 9.7$^\dagger$ & 15.5$^\dagger$ & 8.9$^\dagger$ & 11.6$^\dagger$ & 14.6$^\dagger$\\
33 & GP-TDNN2 &  &  &  &  & 7.6$^\dagger$ & \textbf{14.2}$^\dagger$ & 9.4$^\dagger$ & \textbf{15.1}$^\dagger$ & \textbf{8.7}$^\dagger$ & 11.7$^\dagger$ & \textbf{14.3}$^\dagger$\\
34 & GP-TDNN3 &  &  &  &  & 7.5$^\dagger$ & 14.4$^\dagger$ & \textbf{9.3}$^\dagger$ & \textbf{15.1}$^\dagger$ & 8.9$^\dagger$ & \textbf{11.5}$^\dagger$ & 14.9$^\dagger$\\
35 & V-TDNN &  &  &  &  & 8.0 & 15.3 & 10.1 & 16.3 & 9.3 & 12.6 & 16.0\\
\hline \hline
\end{tabular}
\vspace{-3mm}
\end{table*}

\begin{table*}[h]
\caption{Performance contrasts of LHUC adapted TDNN, B-TDNN, GP-TDNN systems rescored by large RNNLMs against other state-of-the-art systems conducted on the 300-hour Switchboard task. The overall WERs in "()" are not reported by the original papers and are recalculated using the subset WERs.}
\label{swbd900_newest_result}
\centering
\begin{tabular}{c|l|c|ccc|ccc}
\hline\hline
 &
\multirow{2}*{System} &
\multirow{2}*{\#Param} & \multicolumn{3}{c}{Hub5' 00} & \multicolumn{3}{c}{Rt03S}\\
\cline{4-9}
~ & ~ & ~ & SWB1 & CHM & Avg. & FSH & SWB2 & Avg.\\
\hline\hline
1 & RWTH SMBR BLSTM~\cite{kitza2019cumulative} & - & 6.7 & 14.7 & 10.7 & - & - & -\\
2 & + Afﬁne transform based environment adaptation & - & 6.7 & 13.5 & 10.2 & - & - & -\\
\hline\hline
3 & Google Listen, Attend and Spell network + SpecAugment~\cite{park2019specaugment} & - & 6.8 & 14.1 & (10.5) & - & - & -\\
\hline\hline
4 & \multirow{3}*{\shortstack{IBM LSTM based Attention encoder-decoder \\+ SpecAugment + weight noise~\cite{tuske2020single}}} & 29M & 7.4 & 14.6 & (11.0) & - & - & -\\
5 & ~ & 75M & 6.8 & 13.4 & (10.1) & - & - & -\\
6 & ~ & 280M & 6.4 & 12.5 & (9.5) & 8.4 & 14.8 & (11.7)\\
\hline\hline
7 & LF-MMI TDNN + LHUC + Large RNNLM & 19M & 7.5 & 14.5 & 11.0 & 9.5 & 15.5 & 12.6\\
\hline
8 & LF-MMI B-TDNN + LHUC + Large RNNLM &  & 7.0 & 13.9 & 10.5 & 9.0 & 14.4 & 11.8\\
9 & LF-MMI GP-TDNN0 + LHUC + Large RNNLM &  & 7.2 & 14.1 & 10.7 & 9.4 & 14.9 & 12.2\\
10 & LF-MMI GP-TDNN1 + LHUC + Large RNNLM & 19M & 7.3 & 14.1 & 10.7 & 9.1 & 15.0 & 12.2\\
11 & LF-MMI GP-TDNN2 + LHUC + Large RNNLM &  & 7.2 & 13.8 & 10.6 & 9.0 & 14.4 & 11.8\\
12 & LF-MMI GP-TDNN3 + LHUC + Large RNNLM &  & 7.2 & 13.6 & 10.4 & 8.9 & 14.4 & 11.8\\
\hline\hline
\end{tabular}
\vspace{-3mm}
\end{table*}

\begin{enumerate}
    \item Experiments conducted on the Bayesian TDNN systems (B-TDNN, Sec.~\ref{bayesian_est_tdnn}, line 2-6 in Tab.~\ref{swbd75_result}) show that Bayesian estimation only needs to be applied at the first layer, as further applying the Bayesian estimation of any subsequent higher layers produces no additional improvement. This confirms the hypothesis previously discussed in Sec. IV-E that the modelling uncertainty associated with the lower layers of TDNN systems will be much larger than those of the higher layers. Based on this set of experiments, the uncertainty is considered at the first layer of all Bayesian learning based TDNN systems in the rest of the paper.
    \item Compared with the LF-MMI trained TDNN system (line 1 in Tab.~\ref{swbd75_result}), the Bayesian TDNN system (B-TDNN, Sec.~\ref{bayesian_est_tdnn}, line 2 in Tab.~\ref{swbd75_result}) consistently produces a lower WER across all three test sets. For example, the largest absolute WER reduction (1.8\%) is obtained on the SWB5 subset of the Rt02 test set.
    \item The Gaussian Process TDNN systems (GP-TDNN, Sec.~\ref{gp_est_tdnn}, line 8-11 in Tab.~\ref{swbd75_result}) outperformed the Bayesian TDNN system (B-TDNN, Sec.~\ref{bayesian_est_tdnn}, line 2 in Tab.~\ref{swbd75_result}) on the Rt02 test set by 0.2\% (SWB3 subset) to 0.8\% (SWB4 subset) absolute WER reduction, while the Variational TDNN system (V-TDNN, Sec.~\ref{v_est_tdnn}, line 12 in Tab.~\ref{swbd75_result}) was outperformed by the Bayesian TDNN system (B-TDNN, Sec.~\ref{bayesian_est_tdnn}, line 2 in Tab.~\ref{swbd75_result}).
    \item The Bayesian Dropout TDNN system (BD-TDNN, Sec.~\ref{bd_est_tdnn}, line 7 in Tab.~\ref{swbd75_result}) achieves similar performance on three test sets as the Bayesian TDNN system (B-TDNN, Sec.~\ref{bayesian_est_tdnn}, line 2 in Tab.~\ref{swbd75_result}). Based on these results, the Bayesian Dropout TDNN system is not considered in the following large-scale experiments conducted on the 900-hour speed-perturbed Switchboard corpus in Sec.~\ref{exp_swbd900}.
    \item Performance comparison of varying model sizes in Fig.~\ref{fig:model_size_performace_75h} suggests that Bayesian and Gaussian Process TDNN systems are more robust against the change of model sizes, in particular over more complex systems containing up to 250 million parameters (14.7 times of the baseline TDNN system size in line 1 of Tab.~\ref{swbd75_result}). 
\end{enumerate}

\vspace{-3mm}
\subsection{Experiments on 300-Hour Switchboard Task}
\label{exp_swbd900}
To fully evaluate the performance of the Bayesian estimated LF-MMI TDNN systems, experiments were further conducted on the 300-hour (900 hour after speed perturbation) Switchboard conversational English telephone speech recognition task.

{\bf Task Description:} The Switchboard I telephone speech corpus consists of approximately 300 hours audio data released by LDC (LDC97S62). The baseline GMM-HMM system with 6008 tied tri-phone states was trained based on 40-dimensional Mel-frequency cepstral coefficients (MFCCs) to generate alignments for the neural network training. The performance of LF-MMI trained TDNN baseline system incorporated with i-Vector \cite{dehak2011front} and speed perturbation was shown in line 1 of Tab.~\ref{swbd900_result}. In addition, the effects of LHUC~\cite{swietojanski2014learning} speaker adaptation were investigated. For performance evaluation, Kaldi recipe LSTM recurrent neural network language model (RNNLM) trained on the Switchboard and Fisher transcripts (LDC2004T19, LDC2005T19) was used to rescore the nbest lists produced by the LF-MMI trained systems with a four-gram language model (LM).

{\bf Experimental Results and Analysis:} Three main trends can be found in the results of Tab.~\ref{swbd900_result} and Tab.~\ref{swbd900_newest_result}.
\begin{enumerate}
    \item The Bayesian TDNN system (B-TDNN, Sec.~\ref{bayesian_est_tdnn}, line 9 in Tab.~\ref{swbd900_result}) consistently outperforms the TDNN baseline system (line 8 in Tab.~\ref{swbd900_result}) across all three test sets. For example, 0.9\% absolute WER reduction was achieved on the SWB5 subset of Rt02 test set. When compared with the Bayesian TDNN system (B-TDNN, Sec.~\ref{bayesian_est_tdnn}, line 9 in Tab.~\ref{swbd900_result}), the Gaussian Process TDNN system (GP-TDNN2, Sec.~\ref{gp_est_tdnn}, line 12 in Tab.~\ref{swbd900_result}) produced by up to 0.5\% absolute WER reduction on the SWB3 subset of Rt02 test set. On this task, the varational TDNN (V-TDNN, Sec.~\ref{v_est_tdnn}, line 14 in Tab.~\ref{swbd900_result}) made no significant improvements over the TDNN baseline system (line 8 in Tab.~\ref{swbd900_result}). The hidden output distribution in the variational neural network depends on the input data on a frame-by-frame time varying basis. This may introduce undesired artefacts in the resulting hidden layer outputs that are expected to be more invariant to variability in data. This may in part explain the performance difference between V-TDNN and B-TDNN/GP-TDNN systems consistently found in the experiments of this paper.
    
    \item By further incorporating LHUC or Kaldi recipe LSTM recurrent neural network language model (RNNLM) or both of them, similar performance improvements can still be maintained. Statistically significant WER reductions of 0.3\% (SWB1 subset of Hub5'00 test set) to 1.8\% (SWB5 subset of Rt02 test set) absolute (4\% to 11\% relative) were obtained by the Gaussian Process TDNN system (GP-TDNN2, Sec.~\ref{gp_est_tdnn}, line 33 in Tab.~\ref{swbd900_result}) over the TDNN baseline system (line 29 in Tab.~\ref{swbd900_result}).
    \item The experimental results show that the proposed Bayesian TDNN, Gaussian Process TDNN systems (line 2-6, line 30-34 in Tab.~\ref{swbd900_result}) significantly outperform the TDNN baseline systems (line 1, line 29 in Tab.~\ref{swbd900_result}) with and without speed perturbation. This suggests that the proposed method and data augmentation are mostly complementary and their improvements largely additive. 
    \item In order to further evaluate the best performing Bayesian trained systems in Tab.~\ref{swbd900_result}, the LHUC adapted baseline TDNN (line 29 in Tab.~\ref{swbd900_result}), Bayesian TDNN (B-TDNN,  line 30 in Tab.~\ref{swbd900_result}) and Gaussian Process TDNNs (GP-TDNN, line 31-34 in Tab.~\ref{swbd900_result}) were evaluated using a larger RNNLM. The performance of these systems are shown in Tab.~\ref{swbd900_newest_result}. These are then compared with the state-of-the-art performance obtained on the Switchboard task using the most recent hybrid and end-to-end systems reported in the literature (line 1-6 in Tab.~\ref{swbd900_newest_result}). Two larger LSTM recurrent neural network language models (RNNLMs)\footnote{Dropout operation with 85\% retention was applied to the output nodes of each layer.} performing forward and backward contexts based word prediction respectively with twice the number of LSTM cells (2048) and projection dimensionality (1024) compared with the smaller LSTM RNNLM used in Table III were trained. System (1) and System (2) in Tab.~\ref{swbd900_newest_result} were RWTH BLSTM hybrid systems without and with affine transformation for environment adaptation~\cite{kitza2019cumulative}. System (3) was the Google Listen, Attend and Spell end-to-end system built with SpecAugment~\cite{park2019specaugment}. System (4)-(6) were the IBM LSTM based attention encoder-decoder end-to-end systems built with SpecAugment and weight noise~\cite{tuske2020single}. Competitive performance is achieved by the Bayesian estimated TDNN systems (line 8-12 in Tab.~\ref{swbd900_newest_result}) on the CHM subset of Hub5'00 test set and Rt03S test set when compared with the state-of-the-art systems (line 1-6 in Tab.~\ref{swbd900_newest_result}). A general trend can be observed in Tab.~\ref{swbd900_newest_result} such that our B-TDNN and GP-TDNN systems can produce WERs similar to state-of-the-art end-to-end systems with much fewer parameters. For example, by achieving the similar 13.6\% WER on the CHM subset of the Hub5'00 test set, our GP-TDNN3 system only needs 25\% number of parameters of the IBM system (5) producing a comparable WER 13.4\%. Furthermore, our GP-TDNN3 system achieves a state-of-the-art WER of 11.8\% on the Rt03S test set with 93.2\% parameter size reduction when compared with the IBM system (6).
\end{enumerate}

\vspace{-3mm}
\subsection{Experiments on 150-hour HKUST Task}
\label{exp_hkust450}
The performance of Bayesian estimated TDNN systems are further evaluated on a 150-hour (450 hour after speed perturbation) HKUST conversational Mandarin telephone speech recognition task.

{\bf Task Description:} The HKUST Mandarin Telephone Speech contains 150 hours training data released by LDC (LDC2005S15, LDC2005T32). The development set released in 2014~\cite{liu2015cambridge} was used as the validation set. The NIST Rt03S (LDC2007S10) and 1997 NIST Hub5 Mandarin evaluation set~\cite{zhang2015parameterised} were used to form a 2.7-hour test set. Based on the speaker adaptive training (SAT)~\cite{anastasakos1996compact, anastasakos1997speaker, povey2008fast}, a GMM-HMM baseline system with 4000 tied triphone states was trained\footnote{Following the published Kaldi code at github.com/kaldi-asr/kaldi/egs/hkust/s5/run.sh} with 40-dimensional MFCCs and 3-dimensional pitch features~\cite{ghahremani2014pitch} to generate the alignment for the neural network training. The tri-gram language model trained with the training data transcript (LDC2005T32) was used in decoding. All our recognition results were evaluated based on Character Error Rate (CER).

\begin{table}[h]
\vspace{-3mm}
\caption{Performance (CER\%) comparison of TDNN, B-TDNN and GP-TDNN systems on the HKUST mandarin conversational telephone speech \textbf{Dev}, NIST mandarin \textbf{Rt03S} and \textbf{Eval97} evaluation sets. $^\dagger$ denotes a statistically significant difference is obtained over the TDNN baseline system (line 1, 7). Note that significant different test was not performed on the \textbf{Dev} test set.}
\label{hkust_result}
\centering
\setlength{\tabcolsep}{1.5mm}{
\begin{tabular}{c|c|c|c|c|c|c|c}
\hline\hline
 & System &
I-Vector & \tabincell{c}{Speed\\perturb} &
LHUC  & Dev & Rt03S & Eval97\\
\hline\hline
1 & TDNN & \cmark & \cmark & \xmark  & 24.6 & 37.2 & 37.1\\
\hline
2 & B-TDNN &  &  &  &\textbf{23.6} & \textbf{36.2}$^\dagger$ & \textbf{36.1}$^\dagger$\\
3 & GP-TDNN0 &  &  &  &24.5 & 37.0$^\dagger$ & 36.7$^\dagger$\\
4 & GP-TDNN1 & \cmark & \cmark & \xmark  & 24.0 & 36.4$^\dagger$ & 36.6$^\dagger$\\
5 & GP-TDNN2 &  &  &  & 23.9 & \textbf{36.2}$^\dagger$ & 36.5$^\dagger$\\
6 & GP-TDNN3 &  &  &  & 24.0 & 36.6$^\dagger$ & 36.4$^\dagger$\\
\hline \hline
7 & TDNN & \cmark & \cmark & \cmark  & 24.1 & 36.4 & 36.5\\
\hline
8 & B-TDNN &  &  &  & \textbf{23.3} & 35.4$^\dagger$ & \textbf{35.3}$^\dagger$\\
9 & GP-TDNN0 &  &  &  & 23.5 & 35.9$^\dagger$ & 35.6$^\dagger$\\
10 & GP-TDNN1 & \cmark & \cmark & \cmark & \textbf{23.3} & \textbf{35.1}$^\dagger$ & 35.5$^\dagger$\\
11 & GP-TDNN2 &  &  &  & 23.4 & 35.3$^\dagger$ & 35.5$^\dagger$\\
12 & GP-TDNN3 &  &  &  & 23.6 & 35.6$^\dagger$ & 35.9\\
\hline \hline
\end{tabular}}
\vspace{-3mm}
\end{table}

{\bf Experimental Results and Analysis} Two similar trends observed in Tab.~\ref{swbd900_result} can also be found in Tab.~\ref{hkust_result}.
\begin{enumerate}
    \item Compared with the TDNN baseline system (line 1 in Tab.~\ref{hkust_result}), the Bayesian TDNN system (B-TDNN, Sec.~\ref{bayesian_est_tdnn}, line 2 in Tab.~\ref{swbd900_result}) produced 1\% absolute CER reduction across all three test sets. No additional CER improvement was further obtained on the Gaussian Process TDNN systems (GP-TDNN, line3-6 in Tab.~\ref{hkust_result}, Sec.~\ref{gp_est_tdnn}) over the Bayesian TDNN system (B-TDNN, Sec.~\ref{bayesian_est_tdnn}, line 2 in Tab.~\ref{swbd900_result}).
    \item By further incorporating LHUC based speaker adaptation, similar performance improvements can still be maintained. The best performance was achieved by the B-TDNN system ( Sec.~\ref{bayesian_est_tdnn}, line 8 in Tab.~\ref{hkust_result}) and GP-TDNN1 (Sec.~\ref{gp_est_tdnn}, line 10 in Tab.~\ref{hkust_result}). For example, up to 1.3\% absolute CER reduction was achieved on the Rt03S set by GP-TDNN1 system (Sec.~\ref{gp_est_tdnn}, line 10 in Tab.~\ref{hkust_result}) when compared with TDNN baseline system (line 7 in Tab.~\ref{hkust_result}). 
\end{enumerate}

\vspace{-3mm}
\subsection{Experiments on DementiaBank Pitt elderly speech}
\label{exp_db33}
We further evaluate the performance of Bayesian estimated LF-MMI TDNN systems on a cross domain adaptation task which requires porting a 1000 Hour LibriSpeech corpus trained LF-MMI TDNN system to an elderly speech recognition task based on the DementiaBank Pitt database.

{\bf Task Description:} The DementiaBank Pitt corpus\footnote{https://dementia.talkbank.org/access/English/Pitt.html}~\cite{becker1994natural} contains 33-hour audio data, which was split into 27.16-hour training data and 5.81-hour test data. The training data segmentation refinement was first performed by removing excessive silence at the start and end of each utterance in the DementiaBank Pitt corpus. After silence stripping, the DementiaBank Pitt corpus contains 15.75-hour training data (9.72-hour elderly participant data + 6.03-hour investigator data) and 3.14-hour test data (1.93-hour elderly participant data + 1.21-hour investigator data). A GMM-HMM baseline was trained with 39-dimensional PLP features following the same procedures described in~\ref{exp_swbd75}. A 4-gram language model based on the DementiaBank Pitt transcripts, Switchboard and Fisher transcripts and additional text data of 392.4 millions words from the Gigaword collection released by LDC (LDC2011T07) was used in decoding. More details can be found in~\cite{ye2020dev}. The TDNN baseline system was trained with the DementiaBank Pitt data only and its performance was shown in line 7 of Tab.~\ref{exp_db33}. Speed perturbation was also applied to expand the training data to 59-hour in total. The performance of the TDNN system trained on the augmented 59-hour data was shown in line 10 of Tab.~\ref{exp_db33}. The Librispeech corpus~\cite{panayotov2015librispeech} contains 1000 hours of English read speech. Tab.~\ref{librisp_result} shows the performance of the Librispeech based LF-MMI TDNN system\footnote{Following the setup in github.com/kaldi-asr/kaldi/egs/librispeech/s5/run.sh and github.com/kaldi-asr/kaldi/egs/librispeech/s5/local/chain/run$\_$tdnn.sh}. During domain adaptation, the fine-tuning adapted baseline TDNN systems (line 2,5,8,11 in Tab.~\ref{db_result}) reinitialized the input and the output layer of the the LibriSpeech corpus trained LF-MMI TDNN model, while the Bayesian adapted B-TDNN systems (B-TDNN, Sec.~\ref{bayesian_est_tdnn}, line 3,6,9,12 in Tab.~\ref{db_result}) replaced the first layer of the LibriSpeech corpus trained TDNN model with the Bayesian layer and reinitialized the output layer. The baseline TDNN system was cross domain adapted to the Pitt data using parameter fine-tuning, while the B-TDNN system was Bayesian adapted to the same data. The fine-tuning adapted baseline TDNN system serves as the prior of the Bayesian adapted B-TDNN system.

\begin{table}[h]
\vspace{-3mm}
\caption{Performance (WER\%) of the LF-MMI trained TDNN system on the Librispeech \textbf{Dev} and \textbf{Test} test sets.}
\label{librisp_result}
\centering
\begin{tabular}{c|c|c|cc|cc}
\hline\hline
\multirow{2}*{System} &
\multirow{2}*{I-Vector} & \multirow{2}*{Speed perturb} & 
\multicolumn{2}{c}{Dev} & \multicolumn{2}{c}{Test}\\
\cline{4-7}
~ & ~ & ~ & clean & other & clean  & other\\
\hline\hline
TDNN & \cmark & \cmark & 3.56 & 10.0 & 4.0 & 10.3\\
\hline\hline
\end{tabular}
\vspace{-3mm}
\end{table}

{\bf Experimental Results and Analysis} Performance comparison between the fine-tuning adapted baseline TDNN and Bayesian adapted B-TDNN systems was shown in Tab.~\ref{db_result}. Two main trends can be concluded.
\begin{enumerate}
    \item By using different amounts of training data, the Bayesian adapted B-TDNN systems (B-TDNN, Sec.~\ref{bayesian_est_tdnn}, line 3, 6, 9, 12 in Tab.~\ref{db_result}) consistently outperform the fine-tuning adapted baseline TDNN systems (line 2, 5, 8, 11 in Tab.~\ref{db_result}). The best performance was obtained on the Bayesian adapted B-TDNN system (line 12 in Tab.~\ref{db_result}) using the augmented 59-hour DementiaBank Pitt data. This corresponds to a total 1.1\% absolute WER reduction over the fine-tuning adapted baseline TDNN system (line 11 in Tab.~\ref{db_result}). 
    \item When using the 4-hour subset of the Pitt data for cross adaptation, the largest WER absolute reduction up to 2.5\% was obtained by the Bayesian adapted B-TDNN system (B-TDNN, Sec.~\ref{bayesian_est_tdnn}, line 3 in Tab.~\ref{db_result}) over the fine-tuning adapted baseline TDNN system (line 2 in Tab.~\ref{db_result}). 
\end{enumerate}

\begin{table*}[h]
\vspace{-3mm}
\caption{Performance (WER\%) comparison over the DementiaBank Pitt corpus trained TDNN systems, fine-tuning adapted baseline TDNN systems  and Bayesian adapted B-TDNN systems on the \textbf{Participant} and \textbf{Investigator} test sets by using 4-hour, 8-hour 16-hour subset of the Pitt data or an augmented 55-hour Pitt data set. $^\dagger$ denotes a statistically significant difference is obtained over the TDNN baseline system (line 1, 4, 7, 10).}
\label{db_result}
\centering
\begin{tabular}{c|c|c|c|c|c|ccc}
\hline\hline
\multirow{2}*{Row} & \multirow{2}*{System} &
\multirow{2}*{I-Vector} & \multirow{2}*{Speed perturb} & \multirow{2}*{Domain Adaptation} & \multirow{2}*{Training  Data} &
\multirow{2}*{Participant} & \multirow{2}*{Investigator} & \multirow{2}*{All}\\
~ & ~ & ~ & ~ & ~ & ~ & ~ & ~ & ~\\
\hline\hline
1 & TDNN &  &  & \xmark &  &  62.67 & 28.11 & 47.57\\
2 & TDNN & \xmark & \xmark & \cmark & 4h & 58.52 & 26.45 & 44.51$^\dagger$ \\
3 & B-TDNN &  &  & \cmark & &  \textbf{55.95} & \textbf{24.98} & \textbf{42.02}$^\dagger$\\
\hline\hline
4 & TDNN &  &  & \xmark & & 52.79 & 23.99 & 40.21\\
5 & TDNN & \xmark & \xmark & \cmark & 8h & 51.08 & 22.69 & 38.68$^\dagger$\\
6 & B-TDNN &  & & \cmark & &  \textbf{50.32} & \textbf{22.14} & \textbf{38.00}$^\dagger$\\
\hline\hline
7 & TDNN &  &  & \xmark &  & 47.18 & 21.24 & 35.84\\
8 & TDNN & \xmark & \xmark & \cmark & 16h & 45.71 & 20.15 & 34.54$^\dagger$\\
9 & B-TDNN &  &  & \cmark &  & \textbf{44.80} & \textbf{19.82} & \textbf{33.88}$^\dagger$\\
\hline\hline
10 & TDNN &  &  & \xmark &  & 43.88 & 19.84 & 33.37\\
11 & TDNN & \cmark & \cmark & \cmark & 59h & 44.08 & 19.73 & 33.44\\
12 & B-TDNN &  &  & \cmark &  & \textbf{42.53} & \textbf{19.20} & \textbf{32.33}$^\dagger$\\
\hline\hline
\end{tabular}
\vspace{-3mm}
\end{table*}

\section{Conclusion}
This paper presents a full Bayesian framework to account for model uncertainty in sequence discriminative training of factored TDNN acoustic models. Several Bayesian learning based TDNN variant systems are proposed to model the uncertainty over weight parameters and choices of hidden activation functions, or the hidden layer outputs. Efficient variational inference approaches using a few as one single parameter sample ensures their computational cost in both training and evaluation time comparable to that of the baseline TDNN systems. The dropout technique is reformulated as a special case of Bayesian TDNN systems.

Experiments conducted on a state-of-the-art 900 hour speed perturbed Switchboard corpus suggests the proposed Bayesian TDNN, Gaussain Process TDNN and variational TDNN systems consistently outperform the LF-MMI trained TDNN baseline systems by a statistically significant margin of 0.4\%-1.8\% absolute (5\%-11\% relative) reduction in word error rate over the NIST Hub5’00, RT02 and RT03 test sets. Similar consistent performance improvements were also obtained on a 450 hour (with  speed perturbation) HKUST conversational Mandarin telephone speech recognition task. On a third cross domain adaptation task requiring rapidly porting a 1000 hour LibriSpeech data trained system to a 10 hour Dementia Bank elderly speech corpus, the proposed Bayesian TDNN LF-MMI systems outperformed the baseline TDNN system domain adapted using direct weight fine-tuning by 1.1\% absolute WER reduction. 

The proposed Bayesian learning methods applied to TDNNs benefit from a distinct advantage of the underlying latent variable distributions estimation being fully integrated with the overall system training consistently using the same sequence level MMI error cost function. This is in contrast to many existing regularization techniques employed in state-of-the-art speech recognition systems including, not limited to, Gaussian-based weight noise~\cite{murray1994enhanced, braun2019parameter}, L2 norm~\cite{vesely2013sequence} and model averaging~\cite{DBLP:journals/corr/PoveyZK14}. More specifically, the Gaussian-based weight noise is normally kept fixed and not learnable. L2 norm regularisation may be considered as a special form of maximum a posteriori (MAP) estimation~\cite{DBLP:conf/interspeech/HuangSCLWL15}, which restricts the estimation of weight parameters using a fixed Gaussian prior distribution of zero mean and unit variance.  Model averaging that is currently used as a standard regularization method in the Kaldi toolkit~\cite{povey2016purely} can be viewed as averaging over the weight parameters drawn from an unknown distribution that model the parameter estimates obtained at different training epochs or intervals. 

Experimental results obtained across three task domains suggest among all the techniques presented in this paper, the proposed Bayesian TDNNs and Gaussian Process TDNNs (GP-TDNN2 and GP-TDNN3 variants in particular) consistently outperform the baseline TDNN systems featuring state-of-the-art configurations including multiple regularization methods, data augmentation, speaker adaptation and RNNLM rescoring, and therefore are worth further studying on end-to-end speech recognition systems. 

We would also like to note that Gaussian Process TDNNs have the additional ability of modelling neural architecture uncertainty in terms of the suitable activation functions to be used in TDNNs. This unique advantage of GP-TDNNs is as expected and also precisely one of the strengths traditionally associated with Gaussian Processes that is well known for providing powerful non-parametric modelling and black box optimization, for example, in the context of auto-configured Bayesian neural architecture search~\cite{DBLP:conf/nips/KandasamyNSPX18}.

\section*{Acknowledgment}
This research is supported by Hong Kong Research Grants Council GRF grant No. 14200218, 14200220, Theme based Research Scheme T45-407/19N, Innovation \& Technology Fund grant No. ITS/254/19, PiH/350/20, and Shun Hing Institute of Advanced Engineering grant No. MMT-p1-19.

The authors would like to thank Max W. Y. Lam and Yiming Wang for insightful discussions.

\ifCLASSOPTIONcaptionsoff
  \newpage
\fi


\bibliographystyle{IEEEtran}
\bibliography{IEEEexample}
\end{document}